\documentclass[10pt,conference]{IEEEtran}

\usepackage{hyperref}
\usepackage{algorithmic}
\usepackage[scaled=0.81]{beramono}  
\usepackage{bm}
\usepackage{booktabs}               
\usepackage[figurename=Figure]{caption}
\usepackage[T1]{fontenc}
\usepackage{hyphenat} 
\usepackage{listings}
\usepackage{relsize}
\usepackage{url}
\usepackage{xspace}
\usepackage{graphicx}
\usepackage{grffile}  
\usepackage{subcaption}
\usepackage{multirow}

\usepackage[inline]{enumitem}

\usepackage{tikz}
\usetikzlibrary{shapes,arrows,positioning,calc,fit,intersections,through}
\definecolor{inoutcol}{RGB}{253,174,97}
\definecolor{toolcol}{RGB}{171,221,164}
\definecolor{restorecol}{RGB}{160,160,160}
\definecolor{cred}{RGB}{252,141,89}
\definecolor{cblue}{RGB}{145,191,219}
\definecolor{kindA}{RGB}{55,126,184}
\definecolor{kindB}{RGB}{77,175,74}

\newcommand{\bothValid}{36}
\newcommand{\bothCorrect}{23}

\newcommand{\markdef}[1]{$^\ast$#1}

\usepackage{boxedminipage}
\newenvironment{result}%
{\smallskip
   \noindent
   \let\emph=\textbf
   \begin{boxedminipage}{\columnwidth}\begin{center}\em}%
      {\end{center}\end{boxedminipage}%
   \smallskip
}

\newcommand{\jaid}{{\smaller[1]\textsc{Jaid}}\xspace}

\newcommand{\dfj}{{\smaller[1]\textsc{De\-fects\-4J}}\xspace}
\newcommand{\tech}{{\smaller[1]\textsc{Restore}}\xspace}
\newcommand{\restore}{\tech}
\newcommand{\restorefv}{{\smaller[1]\textsc{Restore-full}}\xspace}

\newcommand{\tool}{\tech}

\newcommand{\susp}{\mathit{su}}
\newcommand{\maxsusp}{\ensuremath{\mathit{SU}}}
\newcommand{\snap}{\sigma}
\usepackage{amsmath}
\usepackage{amssymb}
\DeclareMathOperator*{\mean}{mean}

\newcommand{\code}[1]{\mbox{\lstinline[basicstyle=\ttfamily,mathescape=true]|#1|}}

\newcommand{\J}[1]{\code{#1}}

\usepackage{fontawesome}
\newcommand{\fail}{\text{\color{red}\faClose}}
\newcommand{\pass}{\text{\color{green}\faCheck}}

\newcommand{\nicepar}[1]{\textbf{#1}}

\begin{document}

\title{\textsc{Restore}: Retrospective Fault Localization\\ Enhancing Automated Program Repair}

\author{Tongtong Xu, Liushan Chen, Yu Pei, Tian Zhang, Minxue Pan, Carlo A. Furia%
	\IEEEcompsocitemizethanks{
		\IEEEcompsocthanksitem Tongtong Xu is with both the Department of Computing, The Hong Kong Polytechnic University, and the State Key Laboratory for Novel Software Technology, Nanjing University, China.\protect\\ E-mail: \href{mailto:dz1633014@smail.nju.edu.cn}{dz1633014@smail.nju.edu.cn}.
      \IEEEcompsocthanksitem Liushan Chen and Yu Pei are with the Department of Computing, The Hong Kong Polytechnic University, Hong Kong.\protect\\
			E-mail: $\{$\href{mailto:cslschen@comp.polyu.edu.hk}{cslschen}$, $\href{mailto:csypei@comp.polyu.edu.hk}{csypei}$\}${@comp.polyu.edu.hk}.
      \IEEEcompsocthanksitem Tian Zhang is with the State Key Laboratory for Novel Software Technology of Nanjing University, China.\protect\\
			E-mail: \href{mailto:ztluck@nju.edu.cn}{ztluck@nju.edu.cn}. 
      \IEEEcompsocthanksitem Minxue Pan is with the State Key Laboratory for Novel Software Technology and the Software Institute of Nanjing University, China.\protect\\
         E-mail: \href{mailto:mxp@nju.edu.cn}{mxp@nju.edu.cn}.
		\IEEEcompsocthanksitem Carlo A. Furia is with the Software Institute of USI, Universit\`a della Svizzera italiana, Lugano, Switzerland. \protect\\
			Homepage: \url{https://bugcounting.net/}.}%
	\thanks{Received: revised:}}

\maketitle

\begin{abstract}

Fault localization is a crucial step of automated program repair,
because accurately identifying program locations that are most closely implicated with a fault 
greatly affects the effectiveness of the patching process.
An ideal fault localization technique would provide precise information while requiring moderate computational 
resources---to best support an efficient search for correct fixes.
In contrast, most automated program repair tools use standard fault localization techniques---which
are not tightly integrated with the overall program repair process, 
and hence deliver only subpar efficiency.

In this paper, we present \emph{retrospective fault localization}: 
a novel fault localization technique geared to the requirements of automated program repair.
A key idea of retrospective fault localization is 
to reuse the outcome of failed patch validation to support mutation-based dynamic analysis---providing accurate fault localization information
without incurring onerous computational costs.

We implemented retrospective fault localization in a tool called \restore---based on
the \jaid Java program repair system.
Experiments involving faults from the \dfj standard benchmark 
indicate that retrospective fault localization 
can boost automated program repair:
\restore efficiently explores a large fix space,
delivering state-of-the-art effectiveness 
(41 \dfj bugs correctly fixed, 8 of which no other automated repair tool for Java can fix)
while simultaneously boosting performance
(speedup over 3 compared to \jaid).
Retrospective fault localization is applicable to any automated program repair techniques
that rely on fault localization and dynamic validation of patches.
\end{abstract}

\section{Introduction}
\label{sec:intro}

Automated program repair has the potential to transform programming practice:
by automatically building fixes for bugs in real-world programs,
it can help curb the large amount of resources---in time and effort---that programmers devote to debugging~\cite{cost-of-errors}.
While the first viable techniques tended to produce patches that overfit the few tests typically available for validation~\cite{Qi_2015,Smith2015}, 
automated program repair tools have more recently improved precision 
(see \autoref{sec:related-apr} for a review) to the point where they can 
often produce genuinely correct fixes---equivalent to those a programmer would write.

A crucial ingredient of most repair techniques---and especially of so-called \emph{generate-and-validate} 
approaches~\cite{monperrus2014}---is \emph{fault localization}.
Imitating the debugging process followed by human programmers,
fault localization aims to identify program locations that are implicated with a fault
and where a patch should be applied.
Fault localization in program repair has to satisfy two apparently conflicting requirements:
it should be accurate (leading to few locations highly suspicious of error),
but also efficient (not taking too much running time).

In this paper, we propose a novel fault localization approach---called \emph{retrospective fault localization}, and presented in \autoref{sec:details}---that 
improves accuracy while simultaneously boosting efficiency 
by \emph{integrating} closely within standard automated program repair techniques.
By providing a more effective fault localization process,
retrospective fault localization expands the space of possible fixes
that can be searched practically.
Retrospective fault localization leverages mutation-based fault localization~\cite{Moon2014,papadakis_metallaxis_2015}
to boost localization accuracy.
Since mutation-based fault localization is notoriously time consuming,
a key idea is to perform it as a \emph{derivative} 
of the usual program repair process.
Precisely, retrospective fault localization introduces a \emph{feedback loop}
that reuses, instead of just discarding them, the candidate fixes that fail validation to enhance the precision of fault localization.
Candidate fixes 
that pass some tests that the original (buggy) program failed are probably closer to being correct,
and hence they are used to refine fault localization so that other similar candidate fixes are more likely to be generated.

We implemented retrospective fault localization in a tool called \restore,
built on top of \jaid~\cite{chen_contract_2017},
a recent generate-and-validate automated program repair tool for Java.
Experiments with real-world 
bugs from the \dfj curated benchmark~\cite{just2014defects4j}
indicate that retrospective fault localization significantly improves 
the overall effectiveness of program repair 
in terms of correct fixes (for 41 faults in \dfj, 8 more than any other automated repair tool for Java at the time of writing)
and boosts its efficiency
(cutting \jaid's running time to a third or less). 
Other measures of performance, discussed in detail in \autoref{sec:evaluation}, 
suggest that retrospective fault localization 
improves the efficiency of automated program repair by supporting accurate fault localization with comparatively moderate resources.

\nicepar{Generality.}
While our prototype implementation is based on the existing tool \jaid,
retrospective fault localization should be applicable to any program repair tools
that use fault localization and rely on validation through testing.
To demonstrate the approach's generality,
we extended SimFix~\cite{JiangXZGC18}---another state-of-the-art automated repair tools for Java---with
retrospective fault localization.
The experimental results comparing SimFix with and without retrospective fault localization (reported in \autoref{sec:rfl-on-others})
indicate that retrospective fault localization is applicable also to different implementations,
where it similarly brings considerable performance improvements without decreasing effectiveness.

\nicepar{Contributions.}
This paper makes the following contributions:
\begin{enumerate}
   \item Retrospective fault localization: a novel fault localization approach
   tailored for automated program repair techniques based on validation;
   
   \item \restore: a prototype implementation of retrospective fault localization,
   demonstrating how retrospective fault localization can work in practice;
   
   \item An experimental evaluation of \tech on real-world faults from \dfj, showing 
     that retrospective fault localization significantly improves the efficiency by boosting effectiveness and, simultaneously, performance.

   \item An implementation of retrospective fault localization atop the SimFix program repair technique, indicating that it is viable to improve also
     other generate-and-validate repair techniques.
\end{enumerate}

\nicepar{Replication.}
A replication package with \restore's implementation and 
all experimental data is publicly available at: \url{http://tiny.cc/9xff3y}.

\section{An Example of \tech in Action}\label{sec:example}
The \emph{Closure Compiler} is an open source tool that optimizes JavaScript programs to achieve faster download and 
execution times.
One of the refactorings it offers---renaming classes so that namespaces are no longer needed---is based on class \J{ProcessClosurePrimitives} whose methods modify 
calls to common namespace manipulation APIs.
In particular, method \J{processRequireCall} 
processes calls to the \J{goog.require} API and determines if they can be removed without changing program behavior.

\begin{figure}[!b]
\begin{lstlisting}[numbers=left, caption={Faulty method \code{processRequireCall} from class 
\code{ProcessClosurePrimitives} in project \emph{Closure Compiler}.}, label=list:example]  
private void processRequireCall(NodeTraversal t, 
         Node n, Node parent) {
   ProvidedName provided = providedNames.get(...);    (*\label{ln:resolveNS}*)
   ...
   if (provided != null) {                            (*\label{ln:faulty}*)
      parent.detachFromParent();                      (*\label{ln:removeCall}*)
      compiler.reportCodeChange();                    (*\label{ln:bookkeeping}*)
   }
}\end{lstlisting}

\begin{lstlisting}[caption={Fix written by tool developers (replacing line~\autoref{ln:faulty} in 
\autoref{list:example}), and also produced by \tech.}, label=list:human_fix]
if (provided != null || requiresLevel.isOn()) {
\end{lstlisting}
\end{figure}

\autoref{list:example} shows part of the method's implementation, which is defective:\footnote{Fault
\emph{Closure113} in \dfj~\cite{just2014defects4j} and \autoref{tab:results}.} according to the tool
documentation, a call to \J{goog.require} should be removed (lines~\ref{ln:removeCall}
and~\ref{ln:bookkeeping}) if
\begin{enumerate*}[label=\emph{(\roman*)}]
\item \label{cond:checked} the required namespace can be resolved successfully (\J{provided != null}), 
\emph{or}
\item \label{cond:notchecked} the tool is configured to remove all the calls to \J{goog.require} unconditionally 
 (\J{requiresLevel.isOn()}).
\end{enumerate*}
But the code in \autoref{list:example} only checks condition \ref{cond:checked} on line~\ref{ln:faulty}, and hence
does not remove unresolvable calls even when condition \ref{cond:notchecked} holds.

Using some of the tests that come with 
\emph{Closure Compiler}'s source code, the \restore tool described in the present paper
produces the fix shown in \autoref{list:human_fix}, which is identical to the one written by \emph{Closure Compiler}'s tool developers---and
completely fixes the bug.
At the time of writing, \restore is the only automated program repair tool capable of correctly fixing this bug\footnote{Nopol was able to produce a valid, but incorrect, fix to the fault~\cite{durieux17pna}.}.

The features of method \code{processRequireCall} and its enclosing class \J{ProcessClosurePrimitives}
contribute to making the bug challenging for generate-and-validate automated repair tools.
First, class and method are relatively large (Class \J{ProcessClosurePrimitives} has 1233 lines and method \J{processRequireCall} has 40 lines),
which is a challenge in and of itself for precise fault localization.
Second, attribute \J{requiresLevel}  is never referenced in the faulty version of \J{processRequireCall}
and is used only once after initialization in the whole class;
thus, expression \J{requiresLevel.isOn()}---which is needed for the fix---is unlikely to
be selected by techniques that look for fixing ``ingredients'' mainly in a fault's context.

\restore's retrospective fault localization is crucial to ensure that the necessary fixing expression
is found in reasonable time: \restore takes around 32 minutes to produce the fix in \autoref{list:human_fix})
and to rank it first in the output.
This indicates that \restore's search for fixes is not only efficient but also effective.

In the rest of the paper we explain how \restore works (\autoref{sec:details}), and demonstrate its
consistent performance improvements on standard benchmarks of real-world bugs (\autoref{sec:evaluation}).

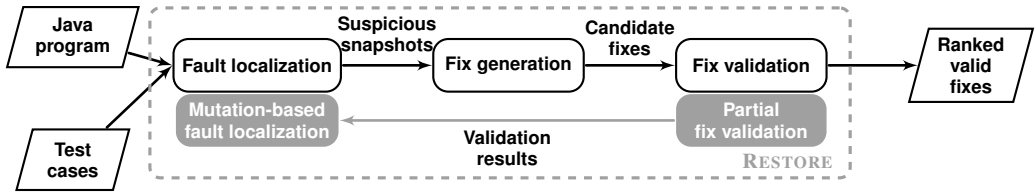
\begin{figure*}[!thb]
   \captionsetup{font=footnotesize}
\centering
\begin{tikzpicture}[
  dbox/.style={rectangle,minimum width=20mm,minimum height=7mm, thick,rounded 
  corners=2mm,fill=white,font=\scriptsize\sffamily\bfseries,text=black, draw=black},
  node distance=8mm and 12mm,
  align=center
  ]
  \node [fill=white,font=\scriptsize\sffamily\bfseries,text=black, draw=black, thick, trapezium,trapezium left angle=75,trapezium right angle=-75] (program) [minimum height=8mm,minimum width=15mm] {Java\\program};
  \node [fill=white,font=\scriptsize\sffamily\bfseries,text=black, draw=black, thick, trapezium,trapezium left angle=75,trapezium right angle=-75] (tests) [below=of program,minimum height=8mm,minimum width=15mm] {Test\\cases};
  \node (mid) at ($(program)!0.23!(tests)$) {};
  \begin{scope}[dbox/.append style={fill=white,text=black, draw=black}]
    \node (localization) [dbox,right=12mm of mid] {Fault localization};
    \begin{scope}
    \node (generation) [dbox,right=of localization] {Fix generation};
    \node (validation) [dbox,right=of generation] {Fix validation};
    \end{scope}
    \begin{scope}[node distance=-1pt,dbox/.append style={fill=restorecol,text=white, draw=none}]
    \node (localization-2) [dbox,below=0.05mm of localization] {Mutation-based\\fault localization};
    \node (validation-2) [dbox,below=0.05mm of validation] {Partial\\fix validation};
    \end{scope}
  \end{scope}
  \node [fill=white,font=\scriptsize\sffamily\bfseries,text=black, draw=black, thick, trapezium,trapezium left angle=75,trapezium right angle=-75] (output) [right=of validation,minimum width=5mm] {Ranked\\valid\\fixes};

  \begin{scope}[color=black,line width=1pt,round cap-latex',every 
  node/.style={font=\scriptsize\sffamily\bfseries}]
  \draw (program) -- (localization.west);
  \draw (tests) -- (localization.west);
  \draw (localization) -- node (snapshots) [above] {Suspicious\\snapshots} (generation);
  \draw (generation) -- node [above] {Candidate\\fixes} (validation);
  \draw (validation) -- (output);
  \begin{scope}[color=restorecol,line width=1pt]
  \draw (validation-2) -- node [below,black] {Validation\\results} (localization-2);
  \end{scope}
  \end{scope}
  \node [fit=(localization)(generation)(validation)(localization-2)(validation-2),
         draw=restorecol, very thick, dashed, rounded corners,inner xsep=3mm,inner ysep=4mm,label={[shift={(-10ex,3ex)},restorecol]south east:\textbf{\tech}}] {};
\end{tikzpicture}
  \caption{An overview of how \tech works. 
\tech can improve the performance of any generate-and-validate automated program repair tool.
Such a tool inputs a faulty program and some test cases exercising the program.
The first, crucial, step of fixing is \emph{fault localization}, which determines a list of snapshots: program states that are indicative of error;
for each suspicious snapshot, \emph{fix generation} builds a number of candidate fixes of the input program by exploring a limited number of program mutations that may avoid the suspicious states;
\emph{fix validation} reruns the available tests on each candidate built by fix generation;
only candidates that pass all tests are \emph{valid fixes}, which are the tool's output to the user.\\
\tech kicks in during the first run of such a program repair tool, by introducing a feedback loop (in {\color{restorecol}grey}) that improves the effectiveness of fault localization.
\tech performs a \emph{partial fix validation}, whose goal is quickly identifying candidate fixes that fail validation---which are treated as \emph{mutants} of the input program; 
information about how mutants' behavior differ from the input program supports a \emph{mutation-based fault localization} step that sharpens the identification of suspicious snapshots.
As we demonstrate in \autoref{sec:evaluation}, \tech's feedback loop significantly improves effectiveness and efficiency of automated program repair.
}
\label{fig:structure}
\end{figure*}

\section{How \tech Works} \label{sec:details}

Retrospective fault localization is applicable in principle to any generate-and-validate automated program 
repair technique to improve its efficiency.
To make the presentation more concrete, we focus on how retrospective fault localization is applicable on top of the \jaid~\cite{chen_contract_2017} automated program repair tool.
We call the resulting technique, and its supporting tool, \restore.

\subsection{Overview}
\autoref{fig:structure} illustrates how \tech works at a high level, and how it enhances a traditional automated program repair technique by retrospective fault localization (boxes {\color{restorecol}in grey} in \autoref{fig:structure}). 

\nicepar{Input.}
\tech inputs a Java program $P$ (a collection of classes), with a faulty method \J{fixme},
and a set $T$ of test cases exercising $P$;
precisely, tests $T$ are partitioned into \emph{passing} tests $T_\pass$ 
and \emph{failing} tests $T_\fail$.
Since each run of \tech actually only uses tests that exercise \J{fixme}, 
we assume, without loss of generality, that $T$ only includes such tests.

\nicepar{Fault localization}
identifies program locations and states (called \emph{snapshots})
that are indicative of faulty behavior.
According to heuristics based on dynamic and static measures, each snapshot receives a \emph{suspiciousness score}---the higher, the more suspicious;
snapshots ranked according to their suspiciousness score are input to the next step: fix generation.

\nicepar{Fix generation}
builds several modifications of input program $P$ for each snapshot in order of suspiciousness.
The modifications try to mutate $P$'s behavior in a way that avoids reaching the suspicious snapshot's state.
Fix generation's output is a sequence of \emph{candidate fixes} that needs to be validated.

\nicepar{(Full) fix validation}
tests each candidate fix to determine wheth\-er it actually fixes the fault exposed by $T_\fail$.
In traditional automated program repair, fix validation runs all available tests $T$ against each fix candidate, and only outputs candidates that pass all tests---ranked according to the suspiciousness of the snapshots they were derived from.
Hence, fix validation is often the most time-consuming step of traditional automated program repair.
Since it is done downstream from fix generation---as the last step of the whole fixing process---validation 
requires a large number of fix candidates to maximize the chance of finding some valid, possibly correct, 
fixes, which exacerbates the performance problem.

\nicepar{Partial fix validation}
is the lightweight form of validation of candidate fixes used by \tech to support retrospective fault localization.
By only running a subset of the available tests $T$, partial fix validation aims to quickly detect 
\emph{behavioral changes} in some of the candidates with respect to the program $P$ under fix. 

\nicepar{Mutation-based fault localization}
improves the precision and effectiveness of fault localization 
by using \emph{retrospective} information coming from partial validation.
Based on this information, the suspiciousness score of snapshots is revised to become more discriminatory.

\nicepar{Exploring a larger fix space.}
With retrospective fault localization, the top-ranked snapshots have a \emph{higher chance} of leading to \emph{valid fixes} when used in the following phases of the repair technique---and thus to correct fixes ranked high in the overall output.
Conversely, a higher-precision fault localization technique means that \emph{fewer candidates} need to be generated and (fully) validated, leading to an overall faster process.
In turn, \restore's more efficient search of the fix space 
allows it to explore a \emph{larger space} in comparable---often \emph{shorter}---time,
ultimately leading to discovering fixes that are outside \jaid's
fix space.

\subsection{Basic Automated Program Repair}\label{sec:jaid}

This section describes the basic process of automated program repair---as implemented 
in generate-and-validate repair tools such as \jaid and \restore.
Then, \autoref{sec:RFL} presents retrospective fault localization in \restore,
showing how it enhances the basic repair process described here.

\subsubsection{State abstraction: snapshots}
\label{sec:snapshots}

\emph{Snapshots} are fundamental abstractions of a program's runs.
A snapshot is a triple $\langle \ell, e, v \rangle$, 
where $\ell$ is a \emph{location} in the program's control-flow graph,
$e$ is a Boolean expression, and $v$ is a Boolean value (\J{true} or \J{false}).
Intuitively, 
$\langle \ell, e, v \rangle$ records the information that a program's run
reaches location $\ell$ with expression $e$ evaluating to $v$.

\restore builds snapshots by enumerating different Boolean expressions $e$ that refer to program features visible at $\ell$, and by evaluating such expressions 
in all runs of tests $T$.

\subsubsection{Fault localization}
\label{sec:traditional-fault-localization}
Fault localization assigns a \emph{suspiciousness score} $\susp(s)$ 
to each snapshot $s$.
Intuitively, $\susp(s)$ should capture the likelihood that $s$ is the source of failure.

Tools like \jaid use a form of spectrum-based fault localization~\cite{Abreu_2006}, 
which roughly corresponds to giving a higher suspiciousness to $s = \langle \ell, e, v \rangle$ 
the more often $e$ evaluates to $v$ at $\ell$ in runs of failing tests than in runs of passing tests.
In \restore, we call \jaid's fault localization \emph{basic fault localization};
\restore uses it to determine a suspiciousness score $\susp_B(s)$ for each snapshot $s$---bootstrapping the fix generation phase.

More precisely, \jaid applies Wong et al.'s Heuristic~III~\cite{EricWong2010} to classify the suspiciousness of \emph{snapshots} rather than statements---as more commonly done in fault localization.
A snapshot $s$'s suspiciousness combines a static analysis score
(measuring the syntactic similarity of the snapshot expression $e$ and the code around location $\ell$)
and a dynamic score
(measuring the relative frequency with which $e = v$ in a failing rather than in a passing test).
Some recent experiments~\cite{Jaid-extended} indicate that \jaid's effectiveness does not significantly depend
on the details of the spectrum-based fault localization algorithm:
running \jaid using other common algorithms for fault localization (such as Ochiai~\cite{Abreu_2006} or Tarantula~\cite{Jones2005EET})
leads to very similar numbers of valid and correct fixes. 

\subsubsection{Fix generation}
\label{sec:traditional-fix-generation}

For each snapshot $\langle \ell, e, v \rangle$, 
fix generation modifies $P$'s method \J{fixme} (the one being fixed)
in ways that affect the value of $e$ at $\ell$.
Fix generation processes snapshots in decreasing order of suspiciousness,
building multiple modifications of \J{fixme} for the same snapshot;
each modification is a \emph{fix candidate}.

\restore generates fix candidates in two steps.
First, it enumerates code snippets (called \emph{actions} in \cite{chen_contract_2017}) that 
\begin{enumerate*}[label=(\alph*)]
\item modify the state of an object referenced in $e$, 
\item modify a subexpression of $e$ in the statement at $\ell$, 
\item if $\ell$ is a conditional statement \J{if (c) ...}, modify expression \J{c}, or
\item modify the control flow at $\ell$ (for example with a \J{return} statement).
\end{enumerate*}
Second, it injects a code snippet \J{action} into \J{fixme} using any of the five schemas in \autoref{fig:schemas}: \J{oldStatement} is the statement at $\ell$ in \J{fixme},
which the whole instantiated schema replaces to generate a fix candidate.

Each fix candidate $C$ can be seen as a mutant of input program $P$ that originates from one snapshot $s$; we write $\snap(C) = s$ to denote the snapshot $s$ that candidate $C$ originates from.
To cull the search space of generated fixes, it is customary to builds fix candidates for at most the top $N$ snapshots in order of suspiciousness; in \jaid, $N = N_S = 1500$.

\begin{figure}[!htb]
   \footnotesize \captionsetup{font=footnotesize}
   \begin{tabular}[B]{l l}
      Schema A: & \J{action; oldStatement;}\\
      Schema B: & \J{if ($e\,$==$\,v$) \{ action; \} oldStatement;}\\
      Schema C: & \J{if ($e\,$!=$\,v$) \{ oldStatement; \}}\\
      Schema D: & \J{if ($e\,$==$\,v$) \{ action; \} else \{ oldStatement; \}}\\
      Schema E: & \J{/* oldStatement; */ action;}\\
   \end{tabular}
   \caption{Schemas to build candidate fixes from a code snippet \J{action} built from snapshot $\langle \ell, e, v \rangle$, where \J{oldStatement} is the statement at $\ell$ in method \J{fixme} under fixing.}
   \label{fig:schemas}
   \normalsize
\end{figure}

\subsubsection{Fix validation (and ranking)}
\label{sec:traditional-fix-validation}

Since fix generation is ``best effort'' and based on the partial information captured by snapshots,
it is followed by a \emph{validation} step that reruns all available tests.
A fix candidate $C$ is \emph{valid} if it passes \emph{all available tests} $T$:
tests $T_\fail$ failing on the input program are passing on $C$, 
and tests $T_\pass$ passing on the input program are still passing on $C$ (no regression errors).

Typically, more than one fix candidate $C$ fixing the same input program $P$ is valid;
we \emph{rank} all such valid fixes in decreasing order of suspiciousness of the snapshot used to generate $C$---that is in decreasing order of $\susp(\snap(C))$.
The overall output of automated program repair is thus a list of valid fixes ranked according to suspiciousness.

\subsection{Retrospective Fault Localization in \tool}
\label{sec:RFL}
The ultimate goal of automated program repair is finding fixes that are not only valid---pass all available tests---but \emph{correct}---equivalent to those a competent programmer, knowledgeable of the program $P$ under repair, would write.
The traditional automated program repair process presented in \autoref{sec:jaid}
can be quite effective at producing correct fixes but is limited in practice by two related requirements: 
\begin{enumerate*}
\item since the accuracy of fault localization greatly affects the
chances of success of the whole repair process, we would like to have a fault localization technique that incorporates as much information as possible;

\item since the process is open loop (no feedback), we have to generate as \emph{many candidate fixes} as possible to maximize the chance of finding a correct one.
\end{enumerate*}
Improving accuracy and generating many candidate fixes 
both exacerbate the already significant problem of \emph{long validation times}
(for example, validation takes up 92.8\% of \jaid's overall running time~\cite{chen_contract_2017}).
More crucially, they require to bound the search space of possible fixes 
to a \emph{size} that can be feasibly explored.
But, by definition, shrinking the fix space makes some bugs impossible to fix.

Retrospective fault localization, as implemented in \restore, addresses these two requirements with complementary solutions:
\begin{enumerate*}
\item it performs a preliminary \emph{partial fix validation}, which runs much faster than full validation and whose primary goal is to supply more dynamic information to fault localization;
\item using the information from partial validation, it complements \jaid's fault localization with precise \emph{mutation-based fault localization}.
\end{enumerate*}
Such a feedback-driven mutation-based fault localization drives more efficient further iterations of fix generation, producing a much smaller, often higher-quality, number of candidate fixes that can undergo full validation taking a reasonable amount of time.
The greater efficiency is then traded off against fix space size:
\restore can afford to explore a \emph{larger space of candidate fixes},
thus ultimately fixing bugs that are out of \jaid's (and other repair tools') capabilities.

\subsubsection{Initial fix generation}
\label{sec:generation-diagnostic}
The initial iteration of fix generation in \restore works similarly to basic automated program repair:
fault localization (\autoref{sec:traditional-fault-localization}) assigns a basic suspiciousness score $\susp_B(s)$ to every snapshot $s$ (using spectrum-based fault localization as in \jaid);
and fix generation (\autoref{sec:traditional-fix-generation}) builds fix candidates for the 
most suspicious snapshots.

As we have already remarked, \jaid's spectrum-based fault localization often takes a major part of the total fixing time, as it involves monitoring the values of many snapshot expressions in every test execution; for example, it takes 51\%--99\% of \jaid's total time on 16 hard faults~\cite{chen_contract_2017}.
To cut down on this major time cost,
\restore \emph{selects} a subset $T_B$ of all tests $T$ to be used in basic fault localization using nearest neighbor queries~\cite{Renieris2003}.
The selected tests $T_B$ include all failing tests $T_\fail$ as well as the passing tests with the \emph{smallest distance} to those failing.
The distance between two tests $t_1, t_2$ is calculated as the Ulam distance\footnote{The Ulam distance~\cite{critchlow1986metric} of two sequences is the minimum number of delete, shift, and insert operations to go from one sequence to another.
  For example, the Ulam distance $U(s_1, s_2)$ of $s_1 = a\,b\,c\,t\,u$ and $s_2 = a\,b\,t\,c\,u$ is $2$ (delete $c$ from $s_1$ and insert it back after $t$).
} $U(\phi(t_1), \phi(t_2))$,
where $\phi(t)$ is a sequence with all basic blocks of \J{fixme}'s control-flow graph sorted according to how many times each block is executed when running $t$. 
This way, passing tests that are behaviorally similar to failing tests are selected as ``more useful'' for fault localization since they are more likely to be sensitive to fixes of the fault.
Take, for example, the conditional at lines~\ref{ln:faulty}--\ref{ln:bookkeeping} in \autoref{list:human_fix};
two tests $t_1$ and $t_2$ such that \J{provided != null} at line~\ref{ln:faulty} both execute the conditional block,
and hence will have a shorter Ulam distance than $t_1$ and another test $t_3$ that skips the conditional block (such that \J{provided == null} at line~\ref{ln:faulty}).
Subset $T_B$ is used only to bootstrap \restore's initial fix generation without dominating the overall running times.

During initial fix generation, \restore builds fix candidates for the $N_1 = N_S \cdot N_P$ most suspicious snapshots (whereas \jaid builds candidates for the $N_S$ most suspicious snapshots).
Parameter $N_P$ is 10\% (i.e., $N_P = 0.1$) by default;
this works because retrospective fault localization can be as effective as \jaid's basic fault localization with a fraction of the snapshots.

\subsubsection{Partial fix validation}
\label{sec:partial-validation}
Partial fix validation aims at quickly extracting dynamic information about 
the many candidate fixes built by the initial iteration of fix generation.
To strike a good balance between costs (time spent on running tests) and benefits (information gathered to guide mutation-based fault localization), 
partial fix validation follows the simple strategy of running only the tests $T_\fail$ 
that were failing on the input program $P$.
This is efficient---because $|T_\fail|$ is often
much smaller than $|T_\pass|$ (see columns \textsc{f} and \textsc{p} in \autoref{tab:results})---and
still has a good chance of providing valuable information for fault localization, 
since it detects whether the failing behavior has changed in some of the fix candidates.

If a candidate fix happens to pass all tests $T_\fail$, 
it immediately undergoes full validation (\autoref{sec:validation-thorough}) for better responsiveness of the fixing process (outputting valid fixes as soon as possible).

\subsubsection{Mutation-based fault localization}
\label{sec:localization-mutationbased}
In mutation-based fault localization~\cite{papadakis_metallaxis_2015,Moon2014}, we compare the dynamic behavior of many different \emph{mutants} of a program.

A mutant is a program variant produced by changing the program's code in some ways---for example, by changing a comparison operator.
A mutant $M$ of a program $P$ is \emph{killed} by a test $t$ when $M$ behaves differently from $P$ on $t$; that is, either $P$ passes $t$ while $M$ fails it, or $P$ fails $t$ while $M$ passes it.
A killed mutant $M$ indicates that the locations where $M$ syntactically differs from $P$ 
are likely (if $M$ fails) or unlikely (if $M$ passes) to be implicated with the failure triggered by $t$.

\restore's retrospective fault localization treats candidate fixes as \emph{higher-order mutants}---that is, mutants of the input program $P$ that may include \emph{multiple} elementary mutations---and interprets partial fix validation results of those higher-order mutants in a similar way to help locate faults more accurately.  
In particular, adapting \cite{papadakis_metallaxis_2015}'s heuristics to our context,
we assign a suspiciousness score $\susp_M(C)$ to each \emph{candidate fix} $C$:
\begin{equation}
  \label{eq:susp-mutation}
  \susp_M(C)\ =\ 
  \frac{| T_\fail \cap \mathit{killed}(C) |}
       {\sqrt{|T_\fail| \cdot |\mathit{killed}(C)|}}\,,
\end{equation}
where $\mathit{killed}(C) \subseteq T_\fail$ is the set of all tests that kill $C$---and 
thus $T_\fail \cap \mathit{killed}(C)$ are the tests that fail on input program $P$ and pass on~$C$.
Formula \eqref{eq:susp-mutation} assigns a higher suspiciousness to a candidate fix the more failing tests it manages to pass, indicating that $C$ might be closer to correctness than $P$.

In order to combine the output of mutation-based and basic fault localization, 
we assign a suspiciousness score $\susp_M(s)$ to each \emph{snapshot} $s$
based on the suspiciousness \eqref{eq:susp-mutation} of \emph{candidates}.
Each candidate fix $D$ is generated from some snapshot $\sigma(D)$;
let $\maxsusp(D)$ be the largest suspiciousness score of all candidate fixes $E$ generated from the same snapshot $\sigma(D)$ as $D$:
\begin{equation*}
  \maxsusp(D) \ =\
\max_{\shortstack{\\$E$}}
\big\{ \susp_M(E) \mid \snap(E) = \snap(D) \big\}\,.
\end{equation*}
Then, the mutation-based suspiciousness score $\susp_M(s)$ of a \emph{snapshot} $s = \langle \ell, e, v \rangle$
is the average of $\maxsusp(D)$ across all candidate fixes $D$ generated from a snapshot with the same location $\ell$ as $s$ (and any expression and value):
\begin{equation}
  \label{eq:susp-mutation-snap}
\susp_M(\langle \ell, e, v \rangle)\ = \ 
\mean_{\shortstack{\\$D$}}
\big\{ \maxsusp(D) \mid \snap(D) = \langle \ell, *, * \rangle \big\}
\,.
\end{equation}
The maximum selects, for each snapshot, the candidate fix generated from it that is more ``successful'' at making failing tests pass.
Then, all snapshots with the same location get the same ``average'' suspiciousness score. 
Intuitively, the average pools the information from different fixes that target different locations and pass partial validation.

Finally, we combine the basic suspiciousness score $\susp_B$ and the mutation-based suspiciousness score $\susp_M$ into an overall total ordering of snapshots according to their suspiciousness:
\[
s_1 \preceq s_2
\ \triangleq\ 
\begin{array}{ll}
     & \big( \ell_1 \neq \ell_2 \ \land\ \susp_M(s_1) \geq \susp_M(s_2) \big) \\
\vee & \big( \ell_1 = \ell_2 \ \land\ \susp_B(s_1) \geq \susp_B(s_2) \big)
\end{array}
\!\!,
\]
where $s_1 = \langle \ell_1, e_1, v_1 \rangle$ and $s_2 = \langle \ell_2, e_2, v_2 \rangle$.
That is, snapshots referring to different locations are compared according to their mutation-based suspiciousness, and snapshots referring to the same location are compared according to their basic suspiciousness---because they have the same mutation-based suspiciousness score.
As discussed in \autoref{sec:traditional-fault-localization}, \restore assigns a basic suspiciousness score to each \emph{snapshot};
whereas the mutation-based suspiciousness score \eqref{eq:susp-mutation-snap} is the same, by definition, for all snapshots with the same location.

\nicepar{An example of how MBFL works.}
To get a more intuitive idea of how mutation-based fault localization can help
find suitable fix locations in \restore, let's consider again
fault \emph{Closure113} in \dfj---shown in Figure~\ref{list:example} and discussed in \autoref{sec:example}.
A single failing test case $T_\fail = \{ t_\fail \}$ triggers the fault by reaching line~\ref{ln:faulty}
with \J{provided == null}: execution skips the \emph{then} branch (lines~\ref{ln:removeCall} and~\ref{ln:bookkeeping}), which eventually leads to a failure.

During the initial round of fix generation,
\restore does not produce any valid fix,
because a key fix ingredient (expression \J{requiresLevel.isOn()})
is further out in the fix search space.
However, it generates 16 candidate fixes 
that happen to pass the originally failing $t_\fail$
because they all force execution through lines~\ref{ln:removeCall} and~\ref{ln:bookkeeping}
by changing condition \J{provided != null} on line~\ref{ln:faulty}.
For example, one such fixes replaces it with
\J{provided != null \|\| provided == null}.
None of these 16 candidates is valid (because they all fail other, previously passing, tests)
but, instead of simply being discarded,
they all are reused as evidence---to
increase the suspiciousness score of line~\ref{ln:faulty}:
\begin{enumerate*}[label=\emph{(\roman*)}]
\item $\susp_M(C) = 1$ for each of these 16 candidates, 
  because $|T_\fail| = 1$ and $\mathit{killed}(C) = T_\fail$;
\item $\maxsusp(C) = \susp_M(C)$ for the same candidates,
  because they all have the same (maximum) value of suspiciousness;
\item $\susp_M(\langle \ell = \ref{ln:faulty}, *, * \rangle) = 1$ for all snapshots that
  target line~\ref{ln:faulty}.
\end{enumerate*}
Since no other candidates generated in this round
change the suspiciousness of other locations,
the net result is that the following iterations of fix generation
will preferentially target fixes at line~\ref{ln:faulty}.
This biases the search for fixes so that \restore goes deeper in this direction
of the fix search space, which eventually leads to generating the correct fix
shown in \autoref{list:human_fix}---which indeed targets line~\ref{ln:faulty}
with a suitable condition.

\subsubsection{Retrospective loop iteration}
\label{sec:branching}
Equipped with the refined fault localization information coming from mutation-based fault localization,
\restore decides whether to iterate the retrospective fault localization loop---entering
a new round of initial fix generation (\autoref{sec:generation-diagnostic})---or
to just use the latest fault localization information to perform a final fix generation (\autoref{sec:generation-concentrated}).
While the retrospective feedback loop could be repeated several times (until all snapshots are used to build candidates),
we found that there are diminishing returns in performing many iterations.
Thus, the default setting is to stop iterating as soon as mutation-based fault localization assigns a \emph{positive} suspiciousness score $\susp_M(s)$ to \emph{some} snapshot $s$;
if no snapshot gets a positive score, we repeat initial fix generation.

\subsubsection{Final fix generation}
\label{sec:generation-concentrated}

Snapshots ranked according to the $\preceq$ relation drive the final generation of fixes.
Final fix generation runs when retrospective fault localization has successfully refined the suspiciousness ranking of snapshots (\autoref{sec:branching})---hopefully identifying few promising snapshots.
Thus, final fix generation generates fixes \emph{only} for snapshots corresponding to the $N_L$ most suspicious locations---with $N_L = 5$ by default.

During final fix generation,
\restore can even afford to trade off some of the greater precision brought by retrospective fault localization for a \emph{larger fix space} to be explored:
whereas \jaid builds fix candidates based only on expressions found in method \J{fixme} (the method being fixed),
\restore may also consider expressions found anywhere in \J{fixme}'s enclosing \emph{class}.
\restore can efficiently search such a larger fix space, thus significantly expanding its overall fixing effectiveness.

\subsubsection{(Full) fix validation}
\label{sec:validation-thorough}
The final validation is, as in basic automated program repair, full---that is, uses \emph{all} available
tests $T$ and validates candidate fixes that pass all of them.
This validation has a higher chance of being significantly faster than in basic automated program repair: 
first, it often has to consider fewer candidate fixes (\autoref{sec:generation-concentrated}) selected according to their mutation-based suspiciousness;
second, several candidate fixes have already undergone partial validation against failing tests $T_\fail$ (\autoref{sec:partial-validation}), and thus only need to be validated against the originally passing tests $T_\pass$.

Fixes that pass validation are output to the user in the same order of suspiciousness $\preceq$ as the snapshots used to generate them.
Thus, \restore's overall output is a list of valid fixes ranked according to suspiciousness.

\section{Experimental Evaluation}
\label{sec:evaluation}

We implemented the \restore technique in a tool, also called \tech, based on the \jaid program repair system. 
Our experimental evaluation assesses to what extent \restore 
is an effective automated program repair tool by comparing:
\begin{enumerate*}[label=\emph{(\roman*)}]
\item \restore's results on high-level metrics, such as \emph{bugs correctly fixed}, to other program repair tools for Java;
\item \restore's results on fine-grained metrics, such as the effectiveness of \emph{fault localization}, to \jaid---a state-of-the-art repair tool for Java which \restore directly extends;
\item the effects of extending SimFix---another recent generate-and-validate repair tool for Java---with \emph{retrospective} fault localization (\restore's key technical improvement).
\end{enumerate*}
Overall, the evaluation indicates that \restore is a substantial advance in general-purpose automated program repair for Java.
Different parts of the evaluation have different levels of granularity,
so that the we can also track \emph{which} ingredients used by \restore are effective and which metrics they impact.

\begin{itemize}[leftmargin=*,labelindent=7mm,labelsep=1.5mm]
   
   \item[\textbf{RQ1:}] What is \tech's \emph{effectiveness} in fixing bugs?
   \\  
   In RQ1, we consider \restore from a user's perspective: 
   how many valid and correct fixes it can generate. 

   \item[\textbf{RQ2:}] What is \tech's \emph{performance} in fixing bugs?
   \\  
   In RQ2, we consider \restore's efficiency: 
   how quickly it runs versus how large a fix space it explores. 

   \item[\textbf{RQ3:}] How well does retrospective \emph{fault localization} (RFL) work in \restore? 
   \\
   In RQ3, we zoom in on \restore's fault localization technique
   to assess how efficiently it drives the search for a valid fix.
   
 \item[\textbf{RQ4:}] How \emph{robust} is \restore's behavior
   when its internal parameters are changed?
   \\
   In RQ4, we evaluate the impact of disabling features like partial validation
   and of changing some parameters that regulate retrospective fault localization.

 \item[\textbf{RQ5:}] Is retrospective fault localization \emph{generally applicable}
   to generate-and-validate program repair techniques?
   \\
   In RQ5, we look for evidence that retrospective fault localization
   is applicable not only to \jaid
   but also to other automated program repair techniques.

\end{itemize}

\nicepar{Comparison to other tools.}
We compare \restore's results on high-level metrics 
to the 13 state-of-the-art automated program
repair systems for Java listed in \autoref{tab:comparison-tools}.
To our knowledge these 13 tools include 
all recent Java repair tools evaluated on \dfj and published, at the time of writing, in major software engineering conferences in the last couple of years.

\subsection{Subject Faults}\label{sec:subjectFaults}
As it has become customary when evaluating automated program repair tools for Java, 
our experiments use real-world faults in the \dfj curated collection~\cite{just2014defects4j}. 
\dfj includes hundreds of faults from open-source Java projects;
each fault comes with at least one test triggering the failure---in addition to other passing or failing tests---as well as a programmer-written fix for the fault.
\autoref{tab:subjects} shows basic measures of size for \dfj's 357 faults in 5 projects.

\begin{table}[!bth]
   \centering
   \setlength{\tabcolsep}{2pt}
   \footnotesize\captionsetup{font=footnotesize}
   \caption{Basic measures of size for projects in \dfj. For each \textsc{project} in \dfj, its \textsc{full name}, the size \textsc{kloc} 
	in thousands of lines of code, the number of tests \textsc{\#tests}, and the number of distinct faults 
	\textsc{\#faults}.}
\label{tab:subjects}
   \begin{tabular*}{\linewidth}{@{\extracolsep{\fill} }llrrr}
      \toprule
      \textsc{project} & \textsc{full name} & \textsc{kloc} & \textsc{\#tests} & \textsc{\#faults} \\
      \midrule
      Chart    & JFreechart         & 96 & 2205 &  26 \\ 
      Closure  & Closure Compiler   & 90 & 7927 & 133 \\ 
      Lang     & Apache Commons-Lang& 22 & 2245 &  65 \\ 
      Math     & Apache Commons-Math& 85 & 3602 & 106 \\ 
      Time     & Joda-Time & 27 & 4130 & 27 \\ 
      \midrule
      & \textsc{total} & 320 & 20109 & 357 \\ 
      \bottomrule
   \end{tabular*}
\end{table}

\subsection{Experimental Protocol}
\label{sec:exp-protocol}

Each experiment runs \tech, \jaid, or another tool to completion on a fault in \dfj.
In each run we record several measures such as:
\begin{itemize}[leftmargin=*,labelindent=6mm,labelsep=1.5mm]
\item[\textsc{\#v}:] number of \emph{valid} fixes in the output;
\item[\textsc{c}:] rank of the first \emph{correct} fix in the output;
\item[\textsc{t}:] overall wall-clock running \emph{time};
\item[\textsc{t2v}:] wall-clock \emph{time} until the first \emph{valid} fix is found;
\item[\textsc{t2c}:] wall-clock \emph{time} until the first \emph{correct} fix is found;
\item[\textsc{c2v}:] number of fixes that are \emph{checked} (generated and validated)
                     until the first \emph{valid} fix is found;
\item[\textsc{c2c}:] number of fixes that are \emph{checked} (generated and validated)
                     until the first \emph{correct} fix is found.
\end{itemize}
Measures \textsc{c2v} and \textsc{c2c} include all kinds of validation.
For example, \restore performs partial and full validation (see \autoref{sec:partial-validation} and \autoref{sec:validation-thorough});
\jaid uses only one kind of (full) validation.
 
\nicepar{Correctness.}
We determined correct fixes by manually going through the output list of valid fixes and comparing each of them to \dfj's manually-written fix for the fault under repair:
a valid fix is correct if it is \emph{semantically equivalent} to the fix manually written by the developers and included in \dfj.
Conservatively, we mark as incorrect fixes that we cannot conclusively establish as equivalent in a moderate amount of time (around 15 minutes per fix).

\nicepar{Hardware/software setup.}
All the experiments ran on the authors' institution's cloud infrastructure. Each experiment used exclusively one virtual machine instance, running Ubuntu 14.04 and Oracle's Java JDK 1.8 on one core of an Intel Xeon Processor E5-2630 v2 with 8~GB of RAM.

\subsubsection{Statistics}
\autoref{tab:summary-results} reports detailed \emph{summary statistics}
directly comparing \restore to \jaid.
For each measure $m$ taken during the experiments (e.g., time \textsc{t}),
let $J_{m, k}$ and $R_{m, k}$ denote the value of $m$ in \jaid's and in \restore's 
run on fault $k$.
We compare \restore to \jaid using these metrics (illustrated and justified below)~\cite{benchmarking}:
\begin{description}
\item[$\frac{\sum \restore}{\sum \jaid}$:] the ratio $\sum_k J_{m, k} / \sum_k R_{m, k}$ 
    expressing the \emph{relative cost} of \restore over \jaid for measure $m$.
  \item[$\mean(\jaid - \restore)$:] the \emph{mean difference} (using arithmetic mean)
    $\mean_k (J_{m, k}$ $- R_{m, k})$  
    expressing the \emph{average additional cost} of \jaid over \restore for measure $m$.
\item [$b_l, \widehat{b}, b_h$:] the estimate $\widehat{b}$ 
    and the 95\% probability interval $(b_l, b_h)$ of the \emph{slope} $b$ of
    the linear regression $R_{m, k} = a + b \cdot J_{m, k}$
    expressing \restore's measure $m$ as a linear function of \jaid's.
\item [$\widehat{\chi}, \chi_h$:] for the same linear regression,
    the estimate $\widehat{\chi}$ and the 95\% probability upper bound $\chi_h$
    of the crossing ratio (where the regression line crosses the ``no effect'' line).
\end{description}
Each summary statistics compares \restore to \jaid on faults on which the statistics is defined for both tools;
for example, the mean difference of measure \textsc{c} (rank of first correct fix) 
is over the \bothCorrect{} faults that \emph{both} \restore and \jaid can correctly fix.

\nicepar{Interpretation of linear regression.}
A linear regression $y = a + b\cdot\! x$ estimates coefficients $a$ (intercept) and $b$ (slope)
in a way that best captures the relation between $x$ and $y$.
A linear regression algorithm outputs \emph{estimates} $\widehat{a}$ and $\widehat{b}$
and \emph{standard errors} $\epsilon_a$ and $\epsilon_b$ for both coefficients:
the ``true'' value of a coefficient $c$ lies in interval $(c_l, c_h)$,
where $c_l = \widehat{c} - 2\,\epsilon_c \leq \widehat{c} \leq \widehat{c} + 2\,\epsilon_c = c_h$,
with 95\% probability.

In our experiments, values of $x$ measure \jaid's performance and values of $y$ measure \restore's;\footnote{In \autoref{sec:rq5}, $x$ measures SimFix's performance and $y$ measures the performance of SimFix+ (SimFix with retrospective fault localization).}
thus, the linear regression line expresses \restore's performance as a linear function of \jaid's.
The line $y = x$ (that is, $a = 0$ and $b = 1$) corresponds to \emph{no effect}: 
the two tool's performances are identical.
In contrast, lines that lie \emph{below} the ``no effect'' line
indicate that \restore measures consistently \emph{lower} than \jaid;
since for all our measures ``lower is better'', 
this means that \restore performs better than \jaid.
Plots such as those in \autoref{fig:linregs} display the estimated {\color{cblue}regression line}
with a shaded area corresponding to the 95\% probability error interval;
thus we can visually inspect whether the difference with respect to the {\color{cred} dashed ``no effect'' line} is significant with 95\% probability by checking whether the shaded area lies under the dashed line.

Analytically, \restore is \emph{significantly better} than \jaid
at the 95\% probability level if the 95\% probability upper bound $b_h$ 
on the regression slope's estimate satisfies $b_h < 1$: the slope
is different from (in fact, less than) the ``no difference'' value $1$ with 95\% probability.

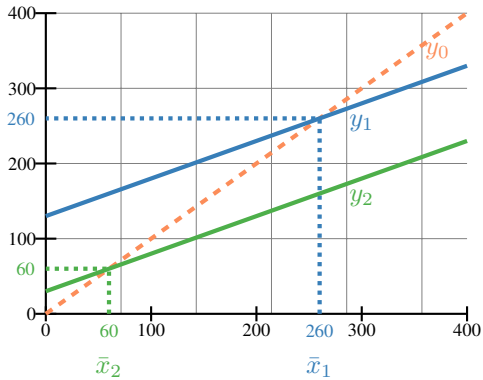
\begin{figure}[!h]
  \centering
  \begin{tikzpicture}[x=14mm]
  \draw[gray,very thin] (0,0) grid (4,4);
  \draw[black,thick] (0,0) -- (4,0);
  \draw[black,thick] (0,0) -- (0,4);
  \foreach \v [count=\x from 0] in {0, 100, 200, 300, 400}
     \draw[thick]  (\x,-0.1) -- node[below] {\scriptsize \v} (\x,0.1);
  \foreach \v [count=\y from 0] in {0, 100, 200, 300, 400}
     \draw[thick]  (-0.1,\y) -- node[left] {\scriptsize \v} (0.1, \y);
     \draw[ultra thick,dashed,cred] (0,0) -- node[very near end,right] {$y_0$} (4,4);
     \draw[ultra thick,kindA] (0,1.3) -- node[near end,below] {$y_1$} (4,3.3);
     \draw[ultra thick,kindA,dotted] (2.6,0) node [below] (A) {\scriptsize 260} -- (2.6,2.6) -- (0, 2.6) node [left] {\scriptsize 260};
     \draw[ultra thick,kindB] (0,0.3) -- node[near end,below] {$y_2$} (4,2.3);
     \draw[ultra thick,kindB,dotted] (0.6,0)  node [below] (B) {\scriptsize 60} -- (0.6,0.6) -- (0,0.6) node [left] {\scriptsize 60};
     \node[below=0pt of A,kindA] {$\bar{x}_1$};
     \node[below=0pt of B,kindB] {$\bar{x}_2$};
\end{tikzpicture}
\caption{Visual explanation of linear regression lines.
  {The two regression lines ${\color{kindA} y_1} = 130 + 0.5\,x$ and ${\color{kindB} y_2} = 30 + 0.5\,y$ have the same slope but different intercepts.
  Therefore, $y_2$ crosses the ``no effect'' line ${\color{cred} y_0} = x$ at $\bar{x}_2 = 60$, much earlier than $y_1$ that crosses it at $\bar{x}_1 = 260$.
  The crossing ratio scales the crossing coordinates $\bar{x}_1$ and $\bar{x}_2$ over the range of values on the $x$ axis.
  If the range is the whole $x$ axis from $0$ to $400$,
  the crossing ratios are simply  $\chi_1 = \bar{x}_1/400 = 0.15$ and $\chi_2 = \bar{x}_2/400 = 0.65$,
  which indicate that $y_1$ is above $y_0$ for only 15\% of the data, and $y_2$ for 65\% of the data. 
}}
\label{fig:cr-example}
\end{figure}

Since this notion of significant difference does not consider the intercept, it only indicates that \restore's is better \emph{asymptotically};
to ensure that the difference is significant in the range of values that were actually measured,
we consider the \emph{crossing ratio} 
$\widehat{\chi} = (\overline{x} - \min(\jaid))/(\max(\jaid) - \min(\jaid))$,
which expresses the coordinate $x = \overline{x}$ where the regression line $y = \widehat{a} + \widehat{b} x$ crosses the ``no effect'' line $y = x$
relative to \jaid's range of measured values
(the crossing ratio upper bound $\chi_h$ is computed similarly but using the upper bounds $a_h$ and $b_h$ of $a$'s and $b$'s 95\% probability 
intervals).
A large crossing ratio means that \restore is better than \jaid only on ``hard'' faults,
whereas a small crossing ratio means that \restore is consistently better across the experimented range, as illustrated in the example of \autoref{fig:cr-example}.

\nicepar{Summarizing data with linear regression.}
Using linear regression to model data that doesn't ``look'' linear may seem unsound.
However, it is not a problem in our case given how we use linear regression:
not to \emph{predict} the performance of \restore on yet to be seen inputs,
but simply to \emph{summarize} the experimental data in a way that accounts for some measurement errors (and hence is more robust than just summarizing the raw data).
After all, the essence of linear regression is a mechanism to
``learn about the mean and variance of some measurement, using an additive combination of other measurements''~\cite{rethinking},
which is all we use it for in analyzing our experimental data.

\subsubsection{Robustness of retrospective fault localization}
\label{sec:describe-robustness}
As described in \autoref{sec:partial-validation},
retrospective fault localization initially performs a \emph{partial} validation
of candidate fixes---using only failing tests.
To understand the usefulness of partial validation, we
built \restorefv: a variant of \restore
that only performs \emph{full} validation---always using all available tests.\footnote{%
  Since full validation may blow up the running time when many tests
  are available for a fault,
  we do not run \restorefv to completion but set a cut-off time
  equal to twice overall running time of \restore on the fault.}
In \autoref{sec:rq4}, we compare \restore and \restorefv on \dfj faults.

In its current implementation, \restore's behavior depends on several parameters:
it uses the $N_S= 1500$ most suspicious state snapshots for fixing (\autoref{sec:traditional-fix-generation});
it adds $N_P=10\%$ more snapshots in each iteration of retrospective fault localization,
and performs $N_I=0$ extra iterations after a new suspicious location has been found (\autoref{sec:RFL});
it targets the $N_L=5$ most suspicious locations for final fix generation (\autoref{sec:generation-concentrated}).
To understand whether these parameters 
influence \restore's behavior,
we modified one of them at a time and ran \restore on the same \dfj faults with these different settings.
In \autoref{sec:rq4}, we report how changing each parameters affects
the number of faults repaired with valid fixes, the number of faults repaired with correct fixes, and the running time across all faults where \restore is able to produce at least one valid fix.

\subsubsection{General application of retrospective fault localization}
\label{sec:rfl-on-others}

To support our claim that retrospective fault localization is applicable
to program repair tools other than \jaid,
we implemented it atop the SimFix~\cite{JiangXZGC18} automated program repair system.\footnote{We used the latest revision \texttt{c2a5319} from SimFix's repository \url{https://github.com/xgdsmileboy/SimFix}.}
We picked SimFix because it is a state-of-the-art repair technique for Java
(as shown in \autoref{tab:comparison-tools},
it correctly fixes the largest number of \dfj bugs when only one fix per bug is considered)
and because its source code and replication package are publicly available.

The key mechanism of retrospective fault localization
is the feedback loop that uses the information gathered during partial validation of candidate fixes to tune fault localization;
this mechanism is general---and hence it is present both in \restore and SimFix+.
On the other hand, \emph{how} the feedback loop collects and processes information, and precisely \emph{when} it does so
depends on the details of the technique to which retrospective fault localization is applied.
Let's see what peculiarities of SimFix affected our implementation of retrospective fault localization in SimFix+.

A key difference between \jaid (and hence \restore) and SimFix
is that the latter's fault localization process,
like most automated repair techniques',
targets \emph{statements} as possible fault locations---rather than snapshots.
Precisely, SimFix applies the Ochiai~\cite{Abreu_2006} spectrum-based fault-localization technique to rank statements according to their suspiciousness.
For each statement above a certain suspiciousness rank,
SimFix searches for ``donor code'' (code snippets in the same project
that are similar to those close to the suspicious statement), 
extracts modification patterns from the donors,
and builds candidate fixes by matching these patterns
to the suspicious statement.
To winnow the many candidate fixes that are generated by this process,
it tries to match them against a ``catalog'' of fixes---which is
generated by mining programmer-written repairs during a preliminary phase
done once before running SimFix on all bugs.
As soon this process determines one fix that is valid (i.e., passes all available tests), SimFix stops.

\begin{table*}[!tbp]
	\small
	\centering
	\caption{A quantitative comparison of \tech with 13 other tools for automated program repair on \dfj bugs. 
		For each program repair \textsc{tool}, the table references
		the source of its experimental evaluation data reported here:
		the number of bugs that the tool could fix with a \textsc{valid} fix;
		the number of bugs that the tool could fix with a \textsc{correct} fix;
		and the resulting \textsc{precision} ($\textsc{correct}/\textsc{valid}$) and \textsc{recall} ($\textsc{correct}/357$, where \textsc{357} is the total number of \dfj faults used in the experiments).
		For tools whose data about the \textsc{position}
		of fixes in the output ranking is available,
		the table breaks down the data
		separately for fixes ranked in \textsc{any position},
		in the \textsc{first positions},
		and in the \textsc{top-10 position}.
		(These measures do not change for tools that output at most one fix per fault.)
		The rightmost column \textsc{unique} lists the number of distinct bugs that \emph{only} the tool can correctly fix.
		Question marks represent data not available for a tool.
	}
	\label{tab:comparison-tools}
	\setlength{\tabcolsep}{2pt}
	\begin{tabular*}{\linewidth}{@{\extracolsep{\fill} }lrrrrrrrrrrr@{}}
		\toprule
		\multirow{2}{*}{\textsc{tool}} & \multirow{2}{*}{\textsc{valid}}  & \multicolumn{3}{c}{\textsc{any position}}    & \multicolumn{3}{c}{\textsc{first position}}  & \multicolumn{3}{c}{\textsc{top-10 position}}  &  \multirow{2}{*}{\textsc{unique}}  \\
		\cmidrule(lr){3-5}\cmidrule(lr){6-8}\cmidrule(lr){9-11}
		& & \textsc{correct}        & \textsc{precision}    & \textsc{recall}       & \textsc{correct}        & \textsc{precision}    & \textsc{recall}  & \textsc{correct}        & \textsc{precision}    & \textsc{recall}    &         \\  \midrule
		\tech & 98    & 41    & 42\%  & 11\%  & 19    & 20\%  & 5\%   & 29    & 30\%  & 8\%   & 8 \\ 
		ACS~\cite{Xiong_2017}   & 23    & 18    & 78\%  & 5\%   & 18    & 78\%  & 5\%   & 18    & 78\%  & 5\%   & 12 \\ 
		CapGen~\cite{WenCWHC18} & 25    & 22    & 88\%  & 6\%   & 21    & 84\%  & 6\%   & 22    & 88\%  & 6\%   & 3 \\ 
		Elixir~\cite{Saha2017} & 41    & 26    & 63\%  & 7\%   & 26    & 63\%  & 7\%   & 26    & 63\%  & 7\%   & 0 \\
		HDA~\cite{xuan_2016}   & ?     & 23    & ?     & 6\%   & 13    & ?     & 4\%   & 23    & ?     & 6\%   & 3 \\
		\jaid~\cite{chen_contract_2017} & 31    & 25    & 81\%  & 7\%   & 9     & 29\%  & 3\%   & 15    & 48\%  & 4\%   & 1 \\
		jGenProg~\cite{DBLP:journals/ese/MartinezDSXM17} & 27    & 5     & 19\%  & 1\%   & 5     & 19\%  & 1\%   & 5     & 19\%  & 1\%   & 1 \\
		jKali~\cite{DBLP:journals/ese/MartinezDSXM17} & 22    & 1     & 5\%   & 0\%   & 1     & 5\%   & 0\%   & 1     & 5\%   & 0\%   & 0 \\
		Nopol~\cite{DBLP:journals/ese/MartinezDSXM17} & 35    & 5     & 14\%  & 1\%   & 5     & 14\%  & 1\%   & 5     & 14\%  & 1\%   & 2 \\
		SimFix~\cite{JiangXZGC18} & 56    & 34    & 61\%  & 10\%  & 34    & 61\%  & 10\%  & 34    & 61\%  & 10\%  & 12 \\
		SketchFix~\cite{HuaZWK18} & 26    & 19    & 73\%  & 5\%   & 9     & 35\%  & 3\%   & ?     & ?     & ?     & 0 \\
		SketchFixPP~\cite{HuaZWK18} & ?     & 34    & ?     & 10\%  & ?     & ?     & ?     & ?     & ?     & ?     & 2 \\
		ssFix~\cite{Xin2017} & 60    & 20    & 33\%  & 6\%   & 20    & 33\%  & 6\%   & 20    & 33\%  & 6\%   & 1 \\
		xPar~\cite{xuan_2016,Xiong_2017}  & ?     & 4     & ?     & 1\%   & ?    & ?     & ?     & 4     & ?     & 1\%   & 0 \\
		\bottomrule
	\end{tabular*}
\end{table*}

We call SimFix+
the modified version of SimFix we built by adding retrospective fault localization.
Just like \restore, SimFix+ undergoes a feedback loop:
after a few candidate fixes are generated,
their partial validation results inform a more accurate iteration of fault localization.
In SimFix+, each iteration of the feedback loop uses $M_P\%$ more code snippets for each suspicious statement to generate a few candidates fixes to ``seed'' retrospective fault localization.
$M_P$ is set to $20\%$ for the initial iterations and $10\%$ for the others, which is usually sufficient to generate enough candidates to drive the process;
if this is not the case (namely, it generates less than 20 candidates), SimFix+ repeatedly increases $M_P$, by $10\%$ each time, until at least 20 candidates are produced or all code snippets are used.

Like in \restore,
partial validation in SimFix+ runs only the \emph{failing} tests for the current bug.
As soon it finds a candidate fix that passes at least one failing test
(``the mutant is killed''), the candidate's fixing location increases its suspiciousness score, and hence SimFix+ immediately begins a new iteration
that generates all fixes at that location and validates them.
This behavior is different from \restore's---where a new iteration only begins after all candidates have undergone partial validation---but is consistent with SimFix's standard behavior of stopping as soon as it finds one valid fix.

In \autoref{sec:rq5}, we experimentally compare SimFix and SimFix+
by running both on \dfj faults.
Each fixing experiment used exclusively one virtual machine instance
running Ubuntu 16.04 on two cores of an Intel Xeon Processor E5-2630 and 8~GB of RAM.
Using the same setting as in the original experiments~\cite{JiangXZGC18},
each SimFix (and SimFix+) run 
is forcefully terminated after a 300-minute timeout if it is still running. 

\subsection{Experimental Results}
\label{sec:results}
\label{sec:whyJaid}

In this section, we report the experiment results as answers to the research questions.

\begin{table*}[!tb]
\centering
\footnotesize\captionsetup{font=small}
\caption{\footnotesize Summary of the experimental results.
	For each fault in \dfj (identified by its \textsc{project} name and \textsc{id})
	that \restore or \jaid can correctly fix:
	the size \textsc{loc} of the faulty method being repaired (in lines of code),
	and the number of \textsc{p}assing and \textsc{f}ailing tests exercising the method;
	for each tool \restore and \jaid:
	the number \textsc{\#v} of \textsc{v}alid fixes;
	the position \textsc{c} of the first \textsc{c}orrect fix in the output;
	the wall-clock running time \textsc{t} to completion;
	the wall-clock running time
	until the first valid fix (\textsc{t2v})
	and the first correct fix (\textsc{t2c}) are found.
	All times are in minutes.}
\label{tab:results}
\begin{tabular*}{\linewidth}{@{\extracolsep{\fill} }lr r r r r r r r r r r r r r r r r r r}
   \toprule
   \multicolumn{2}{c}{\textsc{fault id}}
   & & \multicolumn{2}{c}{\#\textsc{test}}
   & \multicolumn{5}{c}{\textsc{restore}}
   & \multicolumn{5}{c}{\textsc{jaid}}
   \\
   \cmidrule(r){1-2}
   \cmidrule(lr){4-5}
   \cmidrule(lr){6-10}
   \cmidrule(l){11-15}
   \multicolumn{2}{c}{\textsc{project}\hspace{.01cm} \textsc{id}}
   & \multicolumn{1}{c}{\textsc{loc}}
   & \multicolumn{1}{c}{\textsc{p}} 
   & \multicolumn{1}{c}{\textsc{f}}
   & \multicolumn{1}{c}{\textsc{\#v}}
   & \multicolumn{1}{c}{\textsc{c}}
   & \multicolumn{1}{c}{\textsc{t}} 
   & \multicolumn{1}{c}{\textsc{t2v}}
   & \multicolumn{1}{c}{\textsc{t2c}}
   & \multicolumn{1}{c}{\textsc{\#v}}
   & \multicolumn{1}{c}{\textsc{c}}
   & \multicolumn{1}{c}{\textsc{t}} 
   & \multicolumn{1}{c}{\textsc{t2v}}
   & \multicolumn{1}{c}{\textsc{t2c}}
   \\
   \midrule
chart	&	1	&	32	&	37	&	1	&	291	&	221	&	28.5 	&	7.5 	&	21.6 	&	536	&	84	&	54.1	&	5.6	&	19.9	\\
chart	&	9	&	38	&	1	&	1	&	17	&	-	&	14.4 	&	3.3 	&	-	&	52	&	43	&	72.2	&	3.6	&	20.8	\\
chart	&	11	&	32	&	15	&	1	&	1	&	1	&	19.4 	&	17.6 	&	17.6 	&	0	&	-	&	-	&	-	&	-	\\
chart	&	24	&	6	&	0	&	1	&	2	&	1	&	26.7 	&	25.0 	&	25.0 	&	2	&	1	&	16.8	&	15.0	&	15.0	\\
chart	&	26	&	108	&	23	&	22	&	213	&	3	&	32.7 	&	11.5 	&	12.2 	&	82	&	1	&	53.6	&	15.2	&	15.2	\\
closure	&	5	&	98	&	56	&	1	&	4	&	1	&	247.3 	&	186.3 	&	186.3 	&	2	&	-	&	975.9	&	493.5	&	-	\\
closure	&	11	&	18	&	2261	&	2	&	434	&	20	&	846.8 	&	167.5 	&	201.5 	&	0	&	-	&	-	&	-	&	-	\\
closure	&	14	&	97	&	3005	&	3	&	1	&	1	&	355.0 	&	123.5 	&	123.5 	&	0	&	-	&	672.2	&	-	&	-	\\
closure	&	18	&	122	&	3929	&	1	&	1	&	1	&	561.4 	&	101.5 	&	101.5 	&	5	&	1	&	1367.1	&	518.0	&	518.0	\\
closure	&	31	&	122	&	3835	&	1	&	12	&	1	&	570.6 	&	118.4 	&	118.4 	&	9	&	8	&	1440.1	&	1068.2	&	1181.5	\\
closure	&	33	&	27	&	259	&	1	&	171	&	141	&	290.8 	&	19.2 	&	266.7 	&	2720	&	1	&	258	&	6.9	&	6.9	\\
closure	&	40	&	46	&	305	&	2	&	5	&	1	&	25.9 	&	6.1 	&	6.1 	&	4	&	1	&	119.5	&	27.4	&	27.4	\\
closure	&	46	&	11	&	10	&	3	&	161	&	116	&	24.1 	&	4.2 	&	21.3 	&	0	&	-	&	-	&	-	&	-	\\
closure	&	62	&	45	&	45	&	2	&	122	&	90	&	37.5 	&	10.3 	&	30.4 	&	87	&	31	&	126.7	&	8.1	&	31.9	\\
closure	&	63	&	45	&	45	&	2	&	122	&	49	&	34.8 	&	8.8 	&	20.3 	&	87	&	31	&	127.1	&	8.1	&	31.7	\\
closure	&	70	&	19	&	2337	&	5	&	1	&	1	&	127.9 	&	105.3 	&	105.3 	&	5	&	1	&	70.4	&	31.9	&	31.9	\\
closure	&	73	&	70	&	482	&	1	&	1	&	1	&	49.2 	&	39.4 	&	39.4 	&	1	&	1	&	473.4	&	413.5	&	413.5	\\
closure	&	86	&	39	&	52	&	7	&	1	&	1	&	8.9 	&	6.1 	&	6.1 	&	0	&	-	&	-	&	-	&	-	\\
closure	&	113	&	39	&	26	&	1	&	1	&	1	&	48.7 	&	32.5 	&	32.5 	&	0	&	-	&	26.8	&	-	&	-	\\
closure	&	115	&	69	&	151	&	5	&	761	&	1	&	853.4 	&	4.3 	&	4.3 	&	0	&	-	&	-	&	-	&	-	\\
closure	&	118	&	23	&	19	&	2	&	4	&	3	&	33.0 	&	24.6 	&	29.7 	&	0	&	-	&	12.3	&	-	&	-	\\
closure	&	119	&	124	&	764	&	1	&	2	&	2	&	113.5 	&	94.9 	&	113.4 	&	0	&	-	&	-	&	-	&	-	\\
closure	&	125	&	15	&	538	&	1	&	103	&	103	&	154.1 	&	13.1 	&	151.0 	&	98	&	-	&	131.3	&	9.7	&	-	\\
closure	&	126	&	95	&	71	&	2	&	39	&	1	&	103.6 	&	7.8 	&	7.8 	&	425	&	1	&	601.4	&	8.4	&	8.4	\\
closure	&	128	&	9	&	61	&	1	&	14	&	1	&	37.8 	&	9.3 	&	9.3 	&	0	&	-	&	-	&	-	&	-	\\
closure	&	130	&	36	&	301	&	1	&	15	&	4	&	239.1 	&	216.9 	&	221.4 	&	0	&	-	&	-	&	-	&	-	\\
lang	&	6	&	24	&	35	&	1	&	51	&	5	&	142.3 	&	6.6 	&	19.7 	&	0	&	-	&	-	&	-	&	-	\\
lang	&	33	&	11	&	0	&	1	&	3	&	1	&	21.7 	&	11.6 	&	11.6 	&	7	&	1	&	11.0	&	5.5	&	5.5	\\
lang	&	38	&	6	&	33	&	1	&	69	&	18	&	6.7 	&	1.5 	&	4.0 	&	28	&	4	&	10.7	&	1.1	&	1.2	\\
lang	&	45	&	37	&	0	&	1	&	40	&	-	&	35.6 	&	6.5 	&	-	&	68	&	34	&	105.1	&	9.6	&	58.5	\\
lang	&	51	&	51	&	0	&	1	&	37	&	1	&	8.1 	&	4.2 	&	4.2 	&	424	&	46	&	188.4	&	5.4	&	15	\\
lang	&	55	&	6	&	4	&	1	&	29	&	10	&	12.5 	&	1.1 	&	3.0 	&	15	&	3	&	3.6	&	0.4	&	0.9	\\
lang	&	59	&	17	&	2	&	1	&	12	&	7	&	31.7 	&	5.0 	&	11.8 	&	0	&	-	&	-	&	-	&	-	\\
math	&	5	&	22	&	5	&	1	&	225	&	1	&	43.1 	&	3.2 	&	3.2 	&	61	&	1	&	11.3	&	0.6	&	0.6	\\
math	&	32	&	52	&	6	&	1	&	2	&	1	&	10.2 	&	9.2 	&	9.2 	&	5	&	4	&	37.5	&	18.9	&	32.2	\\
math	&	33	&	40	&	21	&	1	&	2	&	2	&	114.9 	&	74.0 	&	74.1 	&	0	&	-	&	251.6	&	-	&	-	\\
math	&	50	&	125	&	3	&	1	&	812	&	94	&	489.2 	&	98.5 	&	137.6 	&	1101	&	28	&	1502.6	&	54.3	&	93.5	\\
math	&	53	&	5	&	19	&	1	&	10	&	9	&	60.0 	&	25.2 	&	51.3 	&	10	&	6	&	19	&	11.1	&	13.3	\\
math	&	59	&	2	&	0	&	1	&	2	&	1	&	3.4 	&	2.4 	&	2.4 	&	0	&	-	&	0.9	&	-	&	-	\\
math	&	80	&	15	&	16	&	1	&	1450	&	936	&	86.9 	&	13.2 	&	65.2 	&	3877	&	1366	&	156.7	&	2.8	&	58.0	\\
math	&	82	&	15	&	13	&	1	&	44	&	22	&	63.9 	&	3.6 	&	25.5 	&	13	&	9	&	33.1	&	3.4	&	22.7	\\
math	&	85	&	43	&	12	&	1	&	235	&	5	&	16.7 	&	3.9 	&	3.9 	&	709	&	4	&	68.3	&	1.5	&	1.5	\\
time	&	19	&	31	&	721	&	1	&	38	&	30	&	15.5 	&	10.4 	&	14.8 	&	0	&	-	&	-	&	-	&	-	\\
\midrule
\multicolumn{2}{r}{\textsc{total}} & 1887	&	19518	&	88	&	5560	&	-	&	6047.1 	&	1645.0 	&	425.9 	&	10433	&	-	&	8998.7	&	2747.7	&	2625.0 \\
   \bottomrule
\end{tabular*}
\end{table*}

\begin{table}[!bht]
\centering
\setlength{\tabcolsep}{1.7pt}
\footnotesize\captionsetup{font=footnotesize}
\caption{\footnotesize 
	Summary statistics of the experiments.
	For each \textsc{measure}:
	the \emph{relative cost} $\frac{\sum \restore}{\sum \jaid}$ of \restore over \jaid;
	the \emph{mean cost difference} $\mean(\jaid - \restore)$ between \jaid and \restore;
	the estimate $\widehat{b}$ of \textsl{slope} $b$ expressing \restore's cost as a linear function of \jaid, with 95\% probability interval $(b_l, b_h)$;
	the estimate $\widehat{\chi}$ and upper bound $\chi_h$ on the \emph{crossing ratio} $\chi$.}
\label{tab:summary-results}
\begin{tabular*}{\linewidth}{@{\extracolsep{\fill} }c c c rrr rr}
   \toprule
   \multirow{2}{*}[-4pt]{\textsc{measure}} & \multirow{2}{*}[-4pt]{$\frac{\sum \restore}{\sum \jaid}$} & \multirow{2}{*}[-4pt]{$\mean(\jaid - \restore)$} & \multicolumn{3}{c}{\textsl{slope} $b$: 95\%}
& \multicolumn{2}{c}{\textsl{crossing} $\chi$}
\\
  \cmidrule(lr){4-6}
  \cmidrule(lr){7-8}
 & 
 & 
 & $b_l$ & $\widehat{b}$ & $b_h$
 & $\widehat{\chi}$ & $\chi_h$
\\
\midrule
\textsc{\#v}  & 0.44 & \makebox[1cm][r]{181} & 0.2 & 0.3 & 0.4 & 0.02 & 0.04
\\
\textsc{c}  & 0.98 & \makebox[1cm][r]{1} & 0.6 & 0.7 & 0.8 & 0.05 & 0.13
\\
\textsc{t}  & 0.32 & \makebox[1cm][r]{214} & 0.2 & 0.2 & 0.3 & 0.02 & 0.04
\\
\textsc{t2v}  & 0.29 & \makebox[1cm][r]{83} & 0.1 & 0.1 & 0.2 & 0.02 & 0.04
\\
\textsc{t2c}  & 0.42 & \makebox[1cm][r]{64} & -0.0 & 0.1 & 0.2 & 0.03 & 0.07
\\
\textsc{c2v}  & 0.43 & \makebox[1cm][r]{1498} & 0.2 & 0.3 & 0.4 & 0.03 & 0.07
\\
\textsc{c2c}  & 0.64 & \makebox[1cm][r]{602} & -0.2 & 0.1 & 0.3 & 0.11 & 0.26
\\

\bottomrule
\end{tabular*}
\end{table}

\subsubsection{RQ1: Effectiveness}
\label{sec:rq1}

RQ1 assesses the \emph{effectiveness} of \restore in terms of 
the \emph{valid} and \emph{correct} fixes it can generate.

Since most automated program repair tools for Java have been evaluated
on the same \dfj bugs as \restore,
we can compare \emph{precision} and \emph{recall}
of the various tools in \autoref{tab:comparison-tools}.\footnote{%
  Since these experimental all refer to the same set of bugs (without cross-validation), precision and recall have a narrower scope as effectiveness metrics here than they have in the context of information retrieval.
} 
\restore and \jaid can output multiple, ranked valid fixes for the same bugs;
in contrast, other tools often stop after producing one valid fix.
We keep this discrepancy into account in \autoref{tab:comparison-tools} by reporting different values of precision and recall according to whether we consider all valid fixes, only those in the top-10 positions, or only those produced in the top position (the first produced).

\nicepar{Valid fixes.}
\restore produced at least one valid fix for 97 faults in \dfj.
As shown in \autoref{tab:comparison-tools},  that is more than any other automated repair tools for Java.

On the \bothValid{} faults that \jaid can also handle,
\restore often produces \emph{fewer valid fixes} than \jaid:
overall, \restore produces 56\% ($1 - 0.44$) fewer valid fixes than \jaid;
and produces more valid fixes for only 13 faults.
As we'll see later, \restore also produces \emph{more} correct fixes than \jaid;
thus, fewer valid fixes per bug can be read as an advantage in these circumstances.

\nicepar{Correct fixes.}
\restore produced at least one correct fix for 41 faults in \dfj---when
considering all fixes for the same bug.
As shown in \autoref{tab:comparison-tools}, that is more than any of the other automated repair tools for Java, and
constitutes a 21\% increase (7 faults) over the runners-up SimFix and SketchFix according to this metric.
\restore correctly fixed 8 faults that \emph{no other tool} can currently fix,
in addition to the 6 faults that only \restore and \jaid can fix. 
This indicates that \restore's fix space is somewhat \emph{complementary} to
other repair tools for Java.

\begin{figure*}[!tb]
	\centering\captionsetup[subfigure]{font=footnotesize}\captionsetup{font=footnotesize}
	\begin{subfigure}[t]{0.2\textwidth}
		\includegraphics[width=\textwidth]{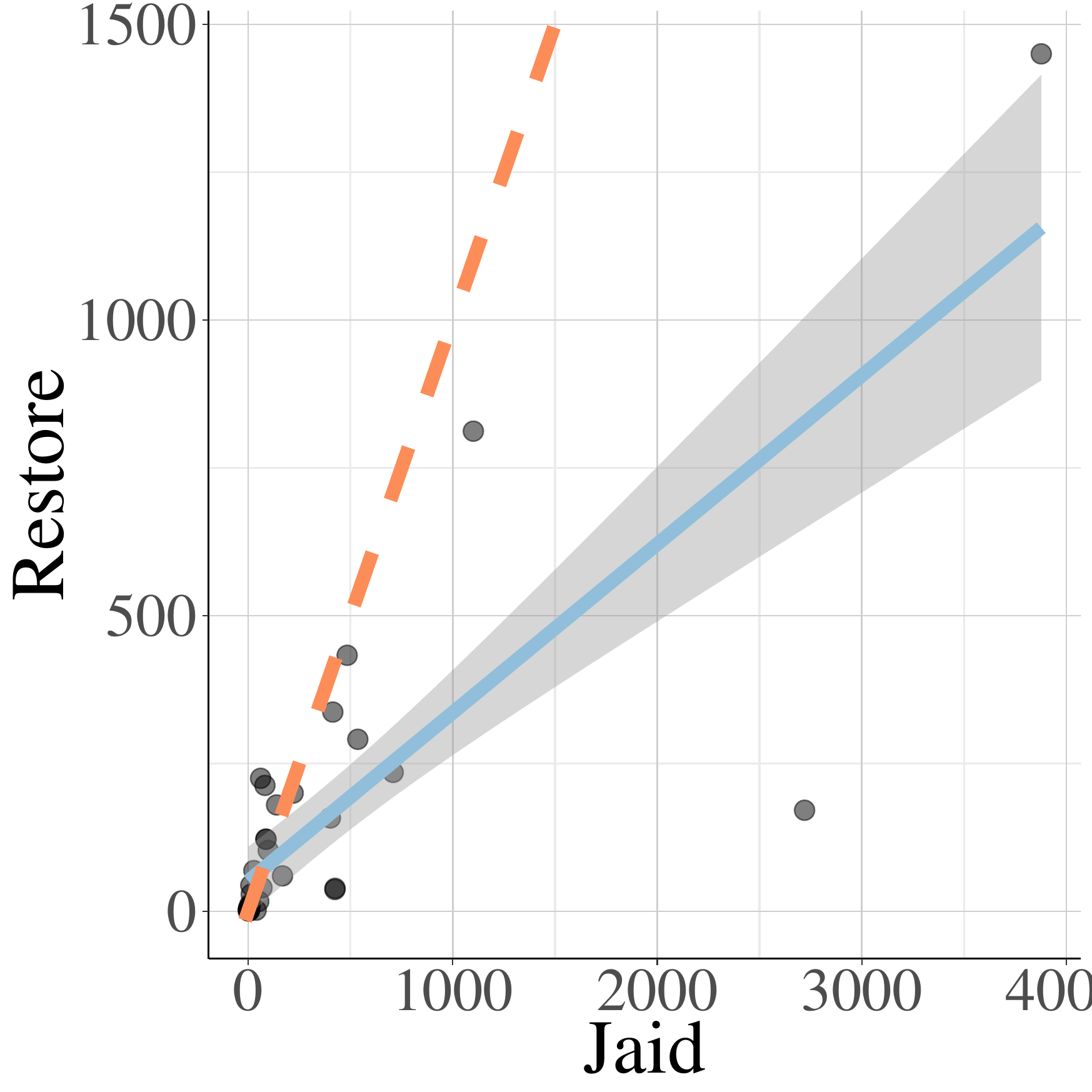}
		\caption{\textsc{\#v}}
		\label{fig:line-v}
	\end{subfigure}  \hfil
	\begin{subfigure}[t]{0.2\textwidth}
		\includegraphics[width=\textwidth]{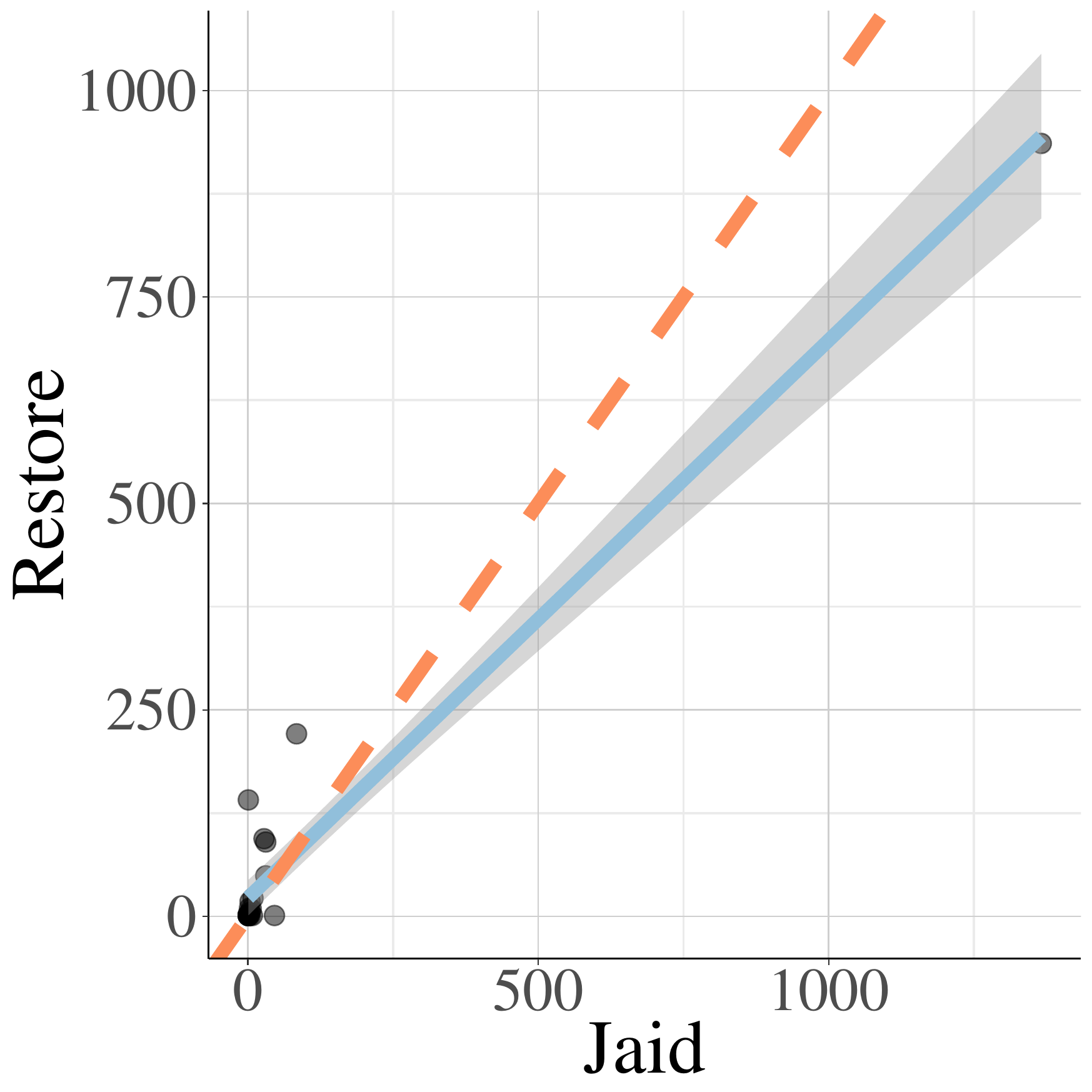}
		\caption{\textsc{c}}
		\label{fig:line-c}
	\end{subfigure}  \hfil
	\begin{subfigure}[t]{0.2\textwidth}
		\includegraphics[width=\textwidth]{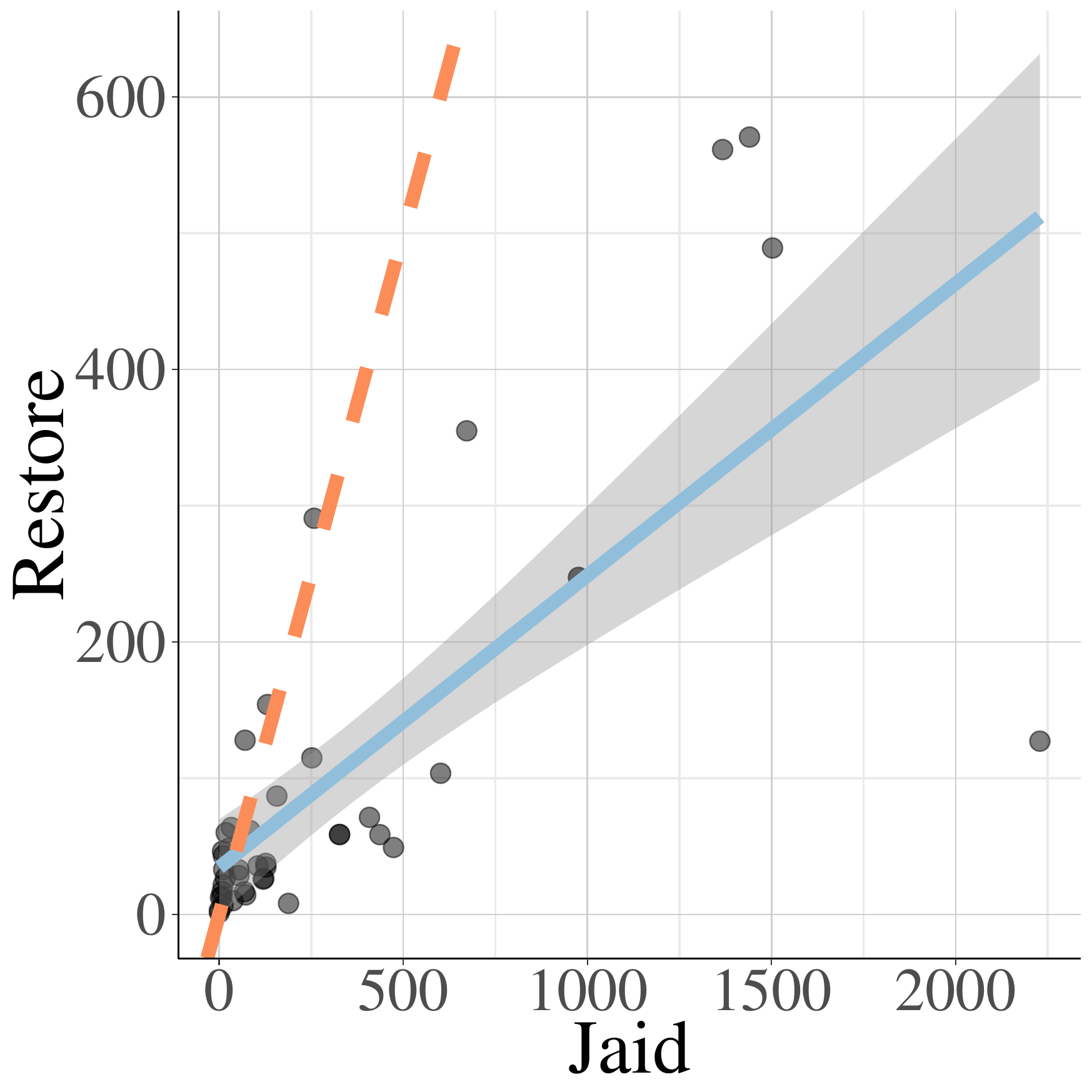}
		\caption{\textsc{t}}
		\label{fig:line-t}
	\end{subfigure}
	
	\begin{subfigure}[t]{0.2\textwidth}
		\includegraphics[width=\textwidth]{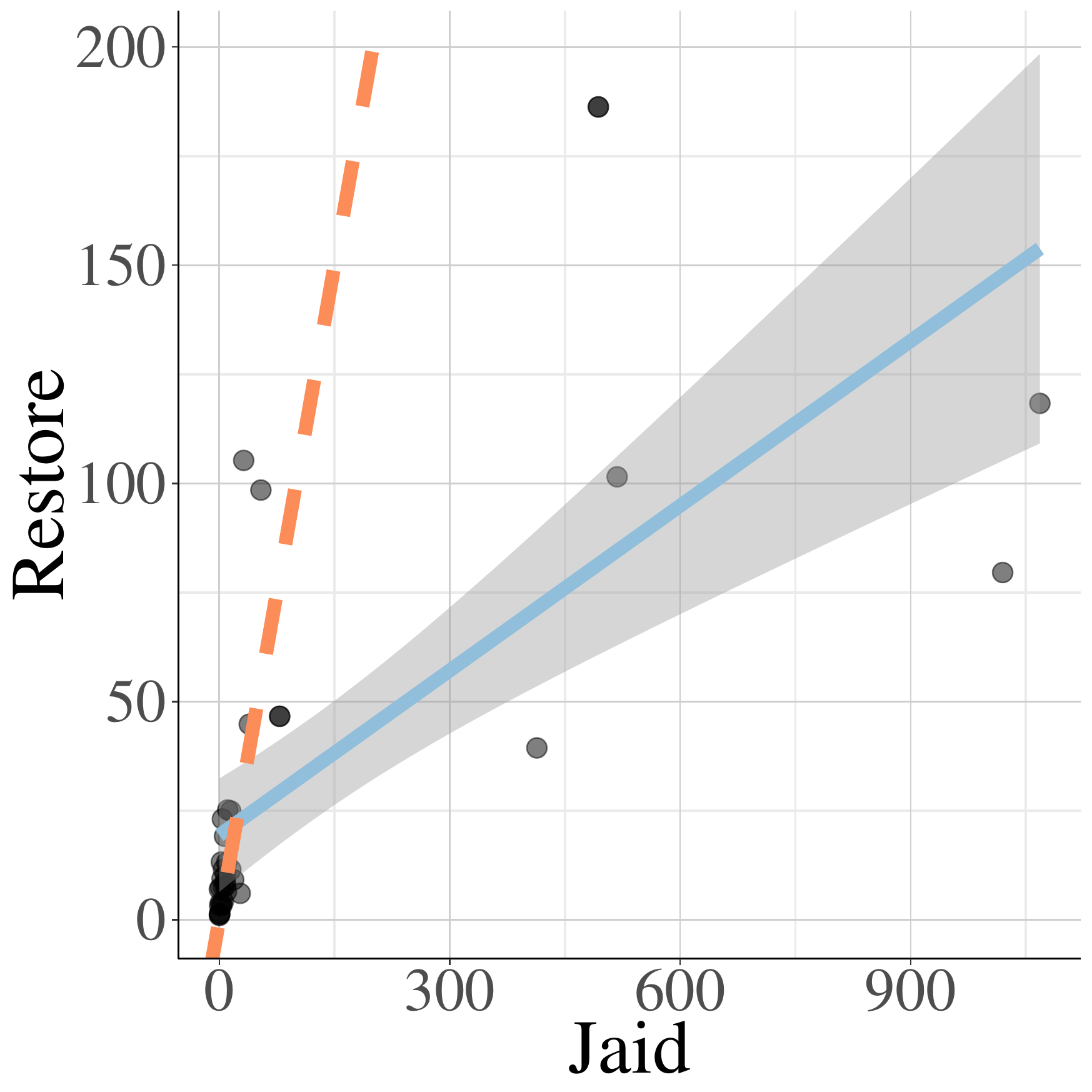}
		\caption{\textsc{t2v}}
		\label{fig:line-t2v}
	\end{subfigure}  \hfil
	\begin{subfigure}[t]{0.2\textwidth}
		\includegraphics[width=\textwidth]{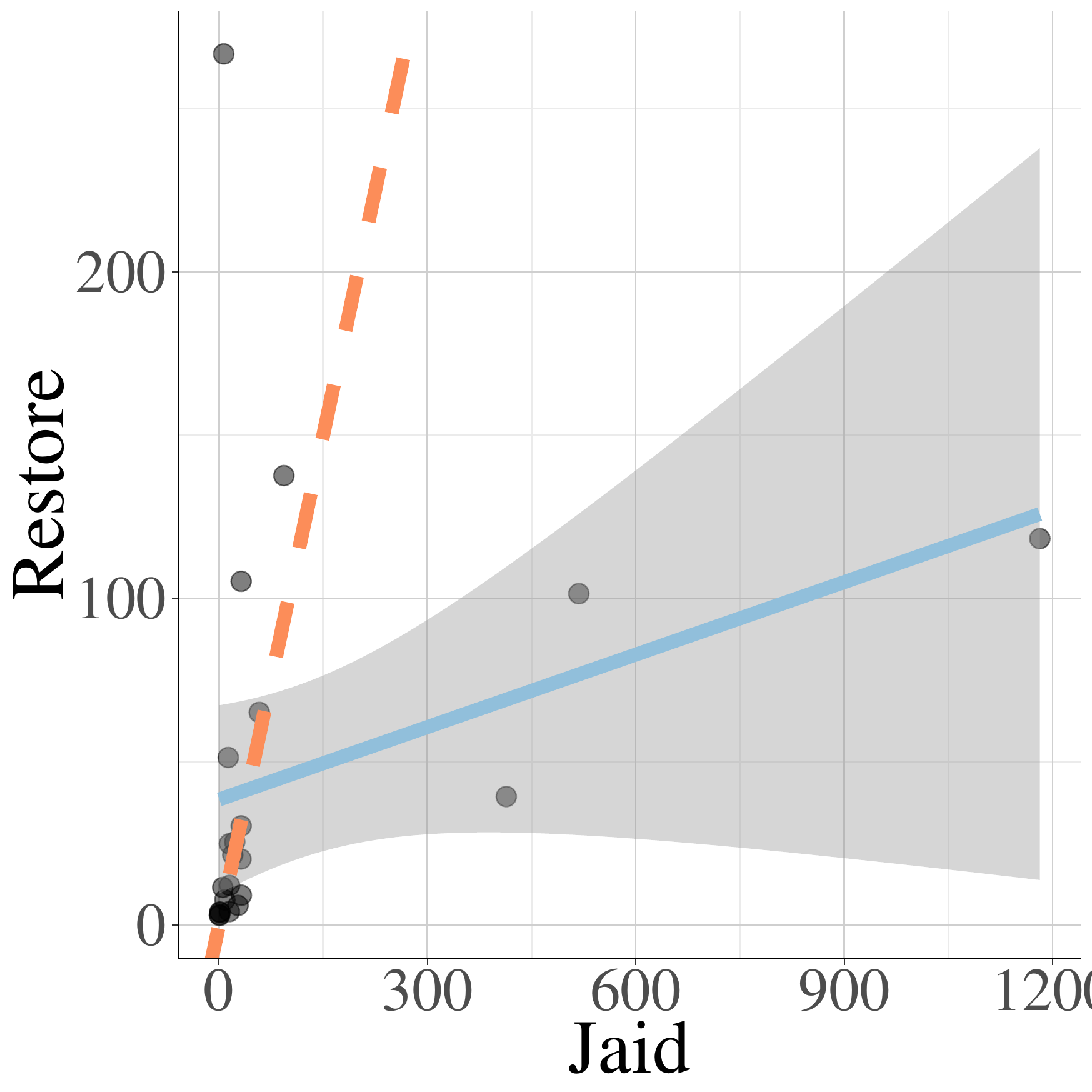}
		\caption{\textsc{t2c}}
		\label{fig:line-t2c}
	\end{subfigure}  \hfil
	\begin{subfigure}[t]{0.2\textwidth}
		\includegraphics[width=\textwidth]{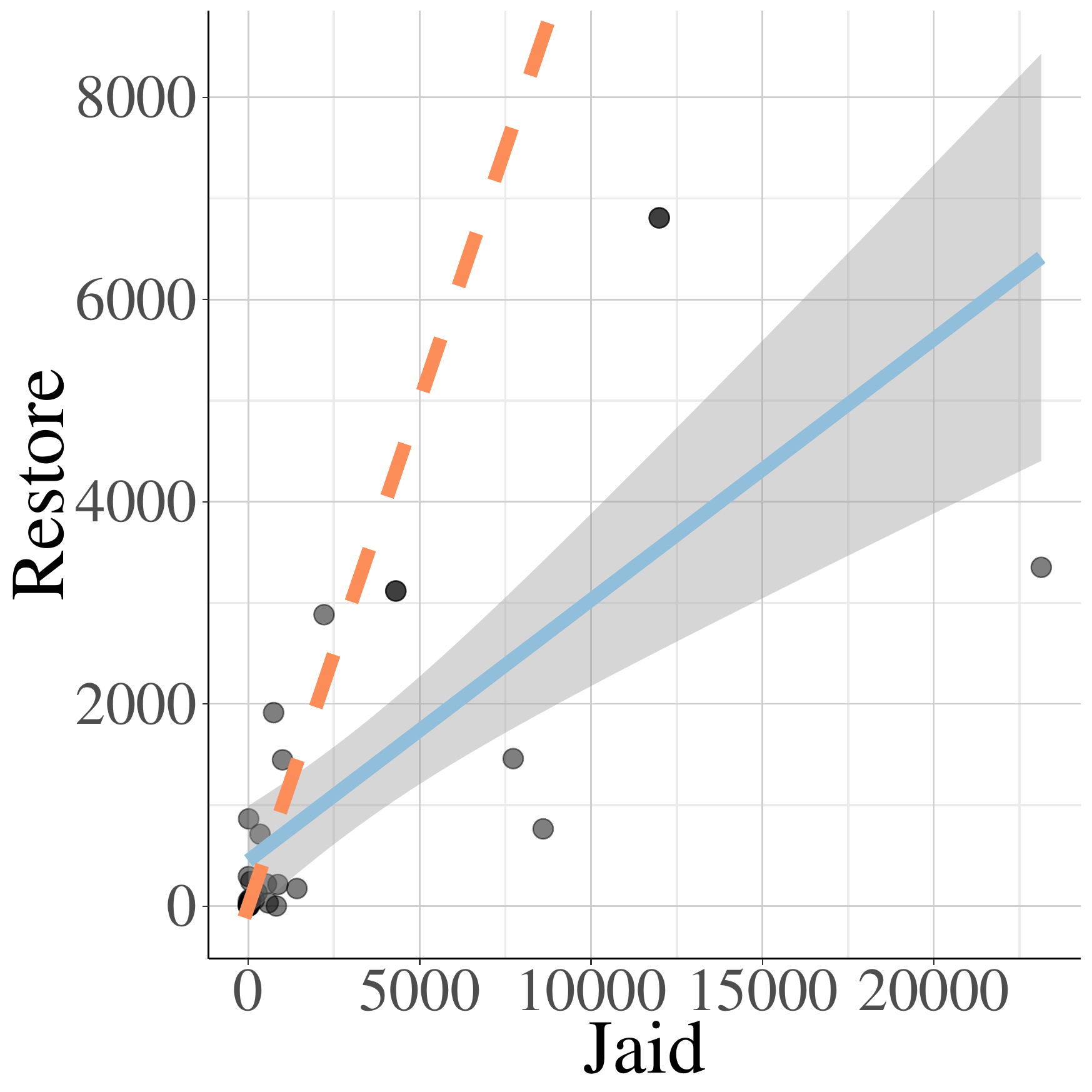}
		\caption{\textsc{c2v}}
		\label{fig:line-c2v}
	\end{subfigure}  \hfil
	\begin{subfigure}[t]{0.2\textwidth}
		\includegraphics[width=\textwidth]{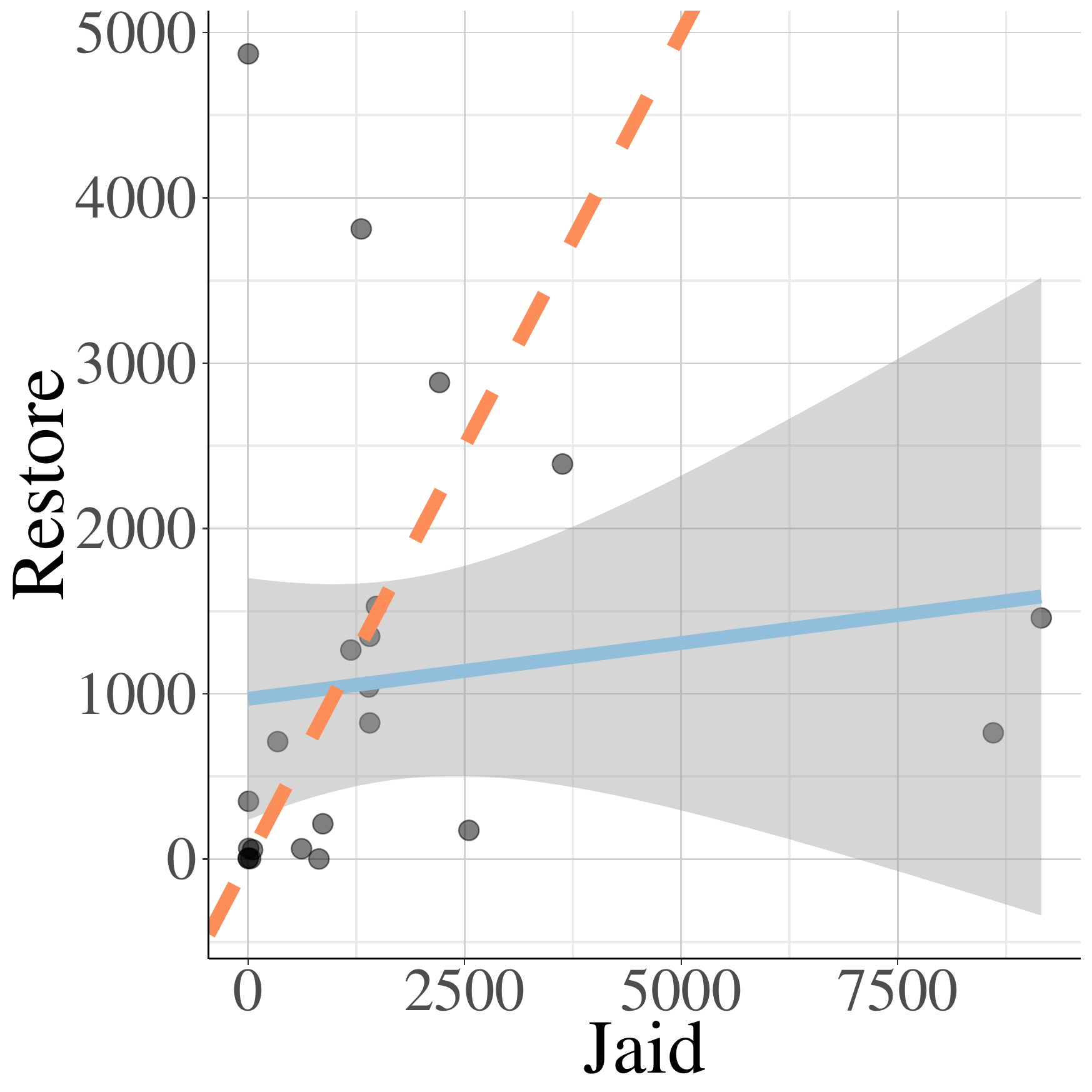}
		\caption{\textsc{c2c}}
		\label{fig:line-c2c}
	\end{subfigure}  
	\caption{Comparison of \jaid and \restore on various measures.
		For each measure $m$, a point with coordinates $x = J_{m, k}, y = R_{m, k}$ 
		indicates that \jaid costed $J_{m, k}$ of $m$ on fault $k$ while \restore costed $R_{m, k}$ of $m$ on fault $k$.
		The {\color{cred}dashed line} is $y = x$; 
		the {\color{cblue}solid line} is the linear regression with $y$ dependent on $x$.}
	\label{fig:linregs}
\end{figure*}

The output list of valid fixes should ideally rank correct fixes \emph{as high as possible}---so
that a user combing through the list would only have to peruse a limited number of fix suggestions.
For the \bothCorrect{} faults that both \restore and \jaid correctly fix,
the two tools behave similarly on the majority of bugs:
\restore ranks the first correct fix 1 position higher than \jaid on average;
and ranks it lower in 11 faults. 
Even thought this difference between the two tools is limited,
\restore still fixes 18 more bugs than \jaid, and ranks first 8 of them.
In addition,
\autoref{fig:line-c} suggests that \restore's advantage over \jaid emerges with ``harder'' faults with many valid fixes---where a reliable ranking is more important for practical usability.

\nicepar{Precision.}
While it can correctly fix more bugs, \restore has a \emph{precision} that is
lower than other repair tools.
In designing \restore we primarily aimed at extending the fix space that can be explored effectively by leveraging retrospective fault localization; 
since there is a trade off between explorable fix space and precision,
the latter is not as high as in other tools that targeted it as a primary goal.

\nicepar{Extended fix space.}
\restore explores a larger fix space than \jaid, since it can also use expressions outside method \J{fixme} in the same class to build fixes
(\autoref{sec:generation-concentrated}).
In all experiments when \restore could produce valid fixes,
68,344 candidate fixes produced during final fix generation
belong to the extended fix space (and hence cannot be produced by \jaid).
Among them, 2,049 candidates are valid (corresponding to 52 faults);
and 9 are correct (one for each of 9 faults).
In all, the extended fix space enabled \restore to generate valid fixes for 17 more bugs than \jaid, correct fixes for 9 more bugs than \jaid;
and correct fixes for 5 of the 8 bugs that only \restore can correctly fix among all tools (\autoref{tab:comparison-tools}).

\nicepar{Multi-line fixes.}
Four of the bugs correctly fixed by \restore (\emph{Closure40}, \emph{Closure46}, \emph{Closure115}, and \emph{Closure128}) have programmer-written fixes in \dfj that change \emph{multiple lines}.
For example, project developers fixed the buggy method of bug \emph{Closure128}:

\begin{lstlisting}[numbers=left,frame=none]
static boolean isSimpleNumber(String s) {
  int len = s.length();
  for (int index = 0; index < len; index++) {
    char c = s.charAt(index);
    if (c < '0' || c > '9') return false; 
  }
  return len > 0 && s.charAt(0) != '0';     
}
\end{lstlisting}
by adding {\small \J{if (len == 0) return false;}} before line 3 \emph{and} changing line~7 to
{\small\J{ return len == 1 \|\| s.charAt(0) != '0';}}.
\restore, instead, just changed line~7 to 
\begin{lstlisting}[frame=none]
  if (len == 1) return true;
           else return len > 0 && s.charAt(0) != '0';
\end{lstlisting}
\restore's conditional return is equivalent to the program-mer-written fix even though it only modifies one location.
Such complex fixes demonstrate how \restore manages to combine bug-fixing effectiveness and competitive performance: this fix was the first valid fix in the output, generated in less than 10 minutes.

\begin{result}
  \restore can correctly fix 41 faults in \dfj when allowing multiple fixes for the same bug; 
  19 of these faults are fixed by the first fix output by \restore.
  \restore trades off a lower precision for a larger fix space,
  which includes correct fixes for 8 faults that no other tools can fix.
\end{result}

\subsubsection{RQ2: Performance}
\label{sec:rq2}
RQ2 assesses the \emph{performance} of \restore in terms of its running time.

\nicepar{Total time.}
\restore's wall-clock total running time per fault ranged between 1.5 minutes and 21 hours,
with a median of 53 minutes.
This means that \restore achieves a speedup of 3.1 ($1/0.32$) over \jaid;
\autoref{fig:line-t} indicates that the major difference in favor of \restore is 
particularly marked for the \emph{harder} faults---which generally require long running times.

Comparing with other tools in terms of running time would require 
to replicate their evaluations using uniform experimental settings---something
we did not do in this experimental evaluation.
Nevertheless, it is plausible other tools have an overall significant running time too:
HDA, ACS, ssFix, Elixir, CapGen, and SimFix are all based on mining external code 
to learn common features of correct fixes;
this process is likely time consuming---even though it would be amortized over a consequent long run of the tools---but is not present in \restore (or \jaid). 
This indicates that \restore's performance is likely to remain competitive overall,
and that retrospective fault localization can bring a performance boon. 
Performing more fine-grained experimental comparisons belongs to future work.

\nicepar{Time to valid/correct.}
Especially important for a repair tool's practical usability 
is the \emph{time elapsing until} a fix appears in the \emph{output}.
All else being equal, shorter times mean that users can start inspecting fix suggestions earlier---possibly supporting a more interactive usage---so that the whole repair process can be sped up.
On average, \restore outputs the first \emph{valid} fix 83 minutes before \jaid---a 3.4 speedup ($1/0.29$) according to the linear regression line;
and the first \emph{correct} fix 64 minutes before \jaid---a 2.3 speedup ($1/0.43$).
While \autoref{fig:line-t2v} and \autoref{fig:line-t2c} suggest that these averages summarize a behavior that varies significantly with some faults, it is clear that \restore's is \emph{substantially faster} in many cases---especially with the ``harder'' faults that require long absolute running times.
Cutting the running times in less than half on average in these cases results in speedups that often span one order of magnitude, and sometimes even two orders of magnitudes.

\restore's performance is the combined result of exploring a larger fix space than \jaid (which takes more time) and using retrospective fault localization (which speeds up fault localization).
That \restore finds many more correct fixes while simultaneously often drastically decreasing the running times indicates that its fault localization techniques bring a decidedly positive impact with no major downsides.

\begin{result}
\restore is usually much faster than \jaid even though it explores a larger fix space:
3.1 speedup in total running time; 3.4 speedup in time to the first valid fix; \\
2.3 speedup in time to the first correct fix.
\end{result}

\subsubsection{RQ3: Fault Localization}
\label{sec:rq3}
\emph{Retrospective fault localization} is \restore's key contribution:
a novel fault localization technique that naturally integrates into generate-and-validate program repair algorithms.
RQ1 and RQ2 ascertained that retrospective fault localization 
indirectly improves program repair by supporting searching 
a larger fix space 
while simultaneously improving performance.
In RQ3 we look into how retrospective fault localization is \emph{directly} more efficient.

\nicepar{Checked to valid/correct.}
To this end, we follow \cite{Qi2013}'s survey of fault localization in automated program repair
and compare the number of fixes that are \emph{checked} (generated and validated) until
the first \emph{valid} (\textsc{c2v}, called NFC in \cite{Qi2013}) and the first \emph{correct} (\textsc{c2c}) fix is generated.
The smaller these measures the more efficiently fault localization drives the search for a valid or correct fix.

\restore needs to check 57\% fewer ($1-0.43$) fixes 
than \jaid
until 
it finds the first valid fix.
\restore significantly improves measure \textsc{c2c} too:
it needs to check 36\% ($1 - 0.64$) fewer fixes than \jaid until 
it finds the first correct fix.
Even though \jaid is more efficient on some faults, 
\autoref{fig:line-c2v} and \autoref{fig:line-c2c} show that \restore prevails in the clear majority of cases, as well as in the harder cases that require to check many more candidate fixes (exploring a larger search space);
the difference is clearly statistically significant (slope under 0.4 with 95\% confidence,
and the overlap of regression line and ``no effect'' line is only for small absolute values of \textsc{c2v} and \textsc{c2c}, as also reflected by the crossing ratio).
These results are direct evidence of retrospective fault localization's
greater precision in searching for fault causes.

\nicepar{Candidate fixes as mutations.}
Retrospective fault localization treats candidate fixes as mutants.
As described in \autoref{sec:localization-mutationbased},
a candidate that passes at least one previously failing test
(during partial validation)
increases the suspiciousness ranking of all snapshots associated
with the candidate's location.
Such candidate fixes sharpen fault localization, and hence we call them \emph{sharpening} candidates.
If a sharpening candidate is furthermore 
associated with a location where a correct fix can be built
(according to the correct fixes actually produced in the experiments or in \dfj)
we call it \emph{plausible}.

\autoref{tab:d-vs-r} measures sharpening and plausible candidates in different categories.
Only 2\% of all candidates are sharpening;
however, the percentage grows to 9\% for faults \restore can build a valid fix for;
and to 12\% for faults \restore can build a correct fix for.
These cases are those where retrospective fault localization achieved progress;
in some cases (\emph{plausible} candidates) it even led
to finding program locations where a correct fix can be built.
\autoref{tab:d-vs-r} also shows that
sharpening and plausible candidates are 9\%
for faults with a single failing test case in \dfj.
These can be considered ``hard'' faults because of the limited information about faulty behavior; retrospective fault localization can perform well even in these conditions.

\autoref{tab:numbersOfIterations} looks at \restore's fault localization feedback loop, which is repeated until retrospective fault localization has successfully refined the suspiciousness ranking.
While some faults require as many as ten iterations,
in most cases only one iteration is needed to achieve progress.
This suggests that candidate fixes are often ``good mutants'' to perform fault localization---and they provide information that is complementary to that available with simpler spectrum-based techniques.

\begin{table}[!bt]
  \centering
  \setlength{\tabcolsep}{2.5pt}
      \caption{How retrospective fault localization achieves progress.
	Each row focuses on faults in one category:
	those that \restore can repair with a \textsc{correct} fix;
	with a \textsc{valid} fix;
	\textsc{all} faults in \dfj;
	and those with a \textsc{single} failing test.
	In each category, the table reports
	how many faults are in total (\textsc{\#});
	for how many \restore's
	fault localization can find a location suitable to build a correct fix
	(\textsc{localized},
	either because \restore actually built a correct fix
	or because the \dfj reference fix modifies that location);
	the number of \textsc{candidates}
	used as mutants in retrospective fault localization;
	how many of these candidates are \textsc{sharpening}
	and \textsc{plausible}.}
\label{tab:d-vs-r}
   \begin{tabular*}{\linewidth}{@{\extracolsep{\fill} }lrrrrrr}
      \toprule
      & \multicolumn{1}{c}{\textsc{\#}} & \multicolumn{1}{c}{\textsc{localized}} & \multicolumn{1}{c}{\textsc{candidates}} & \multicolumn{1}{c}{\textsc{sharpening}} & \multicolumn{1}{c}{\textsc{plausible}}  \\ \midrule 
      \textsc{correct} &  41  &  41  &  23,529   &  2,582  & 511 \\  
      \textsc{valid} & 98  &  75  &  84,989   &  7,348  & 2,762 \\ 
      \textsc{all} & 357 & 107 & 495,359  & 9,854 & 3,377 \\ 
      \textsc{single} & 74 & 57 & 61,530 & 5,307 & 2,108 \\
      \bottomrule
      \end{tabular*}
\end{table}%

\begin{table}[!bt]
	\centering
	\caption{How many times retrospective fault localization
	iterates. Among all faults in \dfj that \restore could repair with a
	\textsc{valid} or a \textsc{correct} fix,
	how many \textsc{iterations} \restore's feedback loop went through
	to sharpen fault localization.}
\label{tab:numbersOfIterations}
	\begin{tabular*}{\linewidth}{@{\extracolsep{\fill} }lrrrrrrrrrr}
     \toprule
     & \multicolumn{10}{c}{\textsc{iterations}} \\
     \cmidrule(lr){2-11}
		& \multicolumn{1}{c}{1} & \multicolumn{1}{c}{2} & \multicolumn{1}{c}{3} & \multicolumn{1}{c}{4} & \multicolumn{1}{c}{5} & \multicolumn{1}{c}{6} & \multicolumn{1}{c}{7} & \multicolumn{1}{c}{8} & \multicolumn{1}{c}{9} & \multicolumn{1}{c}{10} \\
		\midrule 
		\textsc{valid}   & 86 & 3 & 0 & 0 & 3 & 1 & 2 & 0 & 1 & 2  \\
		\textsc{correct} & 35 & 2 & 0 & 0 & 1 & 1 & 1 & 0 & 1 & 0 \\
		\bottomrule
	\end{tabular*}%
\end{table}%

\begin{result}
\restore's retrospective fault localization improves the efficiency of 
the search for correct fixes: on average, 57\% fewer fixes need to be generated and checked until a valid one is found.
The candidate fixes generated by \restore are effective as mutants to perform fault localization.
\end{result}

\subsubsection{RQ4: Robustness}\label{sec:rq4}

RQ4 investigates whether \restore's
overall effectiveness and running time
are affected by changes in features and parameters of its algorithms.

\begin{table}[!bt]
	\centering
	\caption{Comparison between \restore's and \restorefv's effectiveness and performance. The number of \dfj faults with \textsc{valid} fixes,
	with \textsc{correct} fixes,
	and the average running \textsc{time} (in minutes) per fault 
	in \restore compared to those in \restorefv (\restore with only full validation).}
\label{tab:partialOrFullValidation}
	\begin{tabular*}{\linewidth}{@{\extracolsep{\fill} }lrrrr}
      \toprule
       & \textsc{valid} & \textsc{correct} & \textsc{time} \\ \midrule
      \restore  & 98   & 41  & 122.4\\
      \restorefv     & 87   & 27  & 160.6\\
		\bottomrule
	\end{tabular*}%
\end{table}%

\nicepar{Partial validation.} 
\autoref{tab:partialOrFullValidation} summarizes some key performance measures about \restore, and compares them to the same measures for \restorefv---a
variant of \restore that only uses full validation as discussed in \autoref{sec:describe-robustness}.

\restorefv is clearly less effective than \restore,
as the former \emph{misses} valid fixes for 11 faults and correct fixes for 14 faults
that the latter can find.
It is also slower than \restore; in fact, much slower
than what suggested by the 40-minute difference per fault reported in \autoref{tab:partialOrFullValidation}.
Remember that \restorefv is forcefully terminated after it runs for twice as long as \restore on each fault.
With this cap, \restorefv could not complete its analysis for 17 of the 98 faults where \restore produces valid fixes, and it could not even finish the first round of mutation-based fault localization for 13 of them.
(\restore could produce a correct fix for 11 out of these 13 faults.)
Therefore, partial validation is an important ingredient
to make retrospective fault localization scale up, and hence be effective.

\nicepar{Parameters.} 
\autoref{tab:othersettings}
shows how some key performance measures about \restore
change as we individually change the value of
each of four parameters $N_S$, $N_P$, $N_I$, and $N_L$.

The more snapshots $N_S$ are used for fixing,
the more valid and correct fixes \restore can generate.
A closer look indicates a \emph{monotonic} behavior:
if \restore can fix a fault using $s$ snapshots,
it can also fix it using $t > s$ snapshots.
Unsurprisingly, increasing $N_S$ also increases the running time.
Since the number of correctly fixed faults increases only by a few units, whereas the running time increases substantially,
it seems a case of diminishing returns.

\begin{table}[!bt]
	\centering
	\caption{How changing parameters affects \restore's behavior.
	For each \textsc{parameter}
	that control \restore's algorithms,
	the table reports the number
	of \dfj faults with \textsc{valid} fixes,
	with \textsc{correct} fixes,
	and the average running \textsc{time} per fault
	of \restore with different \textsc{value}s of the parameter.
	Values marked with an asterisk (*) are defaults;
	in the experiments where a parameter has a non-default value,
	all other parameters are set to their defaults.}
\label{tab:othersettings}
	\begin{tabular*}{\linewidth}{@{\extracolsep{\fill} }crrrr}
      \toprule
      \textsc{parameter} & \textsc{value} & \multicolumn{1}{c}{\textsc{valid}} & \multicolumn{1}{c}{\textsc{correct}} & \textsc{time} \\ \midrule
      \multirow{3}{*}{$N_S$} &  800  & 90 & 39  & 101.5\\
      & \markdef{1500} &  98 & 41 & 127.0\\
      & 3000 & 103 & 42 & 180.4\\\cmidrule(lr){2-5}
      \multirow{3}{*}{$N_P$} & 5\% & 98 & 39 & 126.6\\
      & \markdef{10\%} & 98 & 39 & 127.0\\
      & 20\% & 99 & 40 & 133.5\\ \cmidrule(lr){2-5}
      \multirow{4}{*}{$N_I$} &  \markdef{0} & 98 & 41 & 127.0 \\
      & 2 & 100 & 41 & 140.4\\
      & 4 & 100 & 40 & 169.1\\
      & 6 & 100 & 41 & 181.6\\\cmidrule(lr){2-5}
      \multirow{3}{*}{$N_L$} & 2 & 91 & 33 & 96.8\\ 
      & \markdef{5}  & 98 & 41 & 124.5 \\
      & 10 & 98 & 41 & 149.9\\ 
		\bottomrule
	\end{tabular*}%
\end{table}

In contrast, the effects of changing the percentage $N_P$ of snapshots used in each iteration of retrospective fault localization are very modest---both on the running time and on the number of valid and correct fixes.
Increasing $N_I$---that is,
iterating retrospective fault localization even after it has contributed to refining the ranking of suspicious locations---also has a modest effect on effectiveness but noticeably increases the running time.
Overall, \restore's behavior is not much affected by how snapshots are sampled,
but repeating retrospective fault localization beyond what is needed
tends to decrease \restore's efficiency without any clear advantage.

The default value of parameter $N_L$---the number of most suspicious locations used for final fix generation (\autoref{sec:generation-concentrated})---seems
to strike a good balance between effectiveness and efficiency:
increasing $N_L$
does not lead to fixing more faults, but visibly increases the running time;
decreasing it reduces the running time, but also fixes fewer faults.

\begin{result}
  Partial validation is crucial for the efficiency
  of retrospective fault localization.
  \restore's effectiveness is usually only weakly dependent on the values of internal parameters.
\end{result}

\subsubsection{RQ5: Generalizability}
\label{sec:rq5}

By comparing SimFix to SimFix+ (our variant of SimFix
that implements retrospective fault localization) RQ5 analyzes
the applicability of retrospective fault localization
to tools other than \restore.

Both SimFix and SimFix+ can build \emph{valid} fixes
for the same 64 faults in \dfj.
SimFix can generate valid fixes for another 4 faults that SimFix+ cannot,
and hence can fix 68 faults in total;
conversely, SimFix+ can generate valid fixes for another 7 faults that SimFix cannot,
and hence can fix 71 in total.
In the case of the 4 faults that only SimFix can repair, SimFix's simple spectrum-based fault localization was sufficiently precise to guide the process to success (by ranking high locations that lead to suitable donor code).
In contrast, the donor code leading to candidates that are useful for mutation-based fault localization (see \autoref{sec:rfl-on-others}) was ranked low;
thus, 
SimFix+'s retrospective fault localization took multiple iterations and a long time to go through the many candidates,
and ended up hitting the tool's 300-minute timeout.
The cases of the 7 faults that only SimFix+ can repair are opposite:
spectrum-based fault localization was
imprecise, hampering the performance of SimFix,
whereas mutation-based fault localization could successfully complete its analysis
and sharpen the suspiciousness ranking as required by these 7 faults.

\begin{figure}[!tb]
	\centering
	\includegraphics[width=0.48\textwidth]{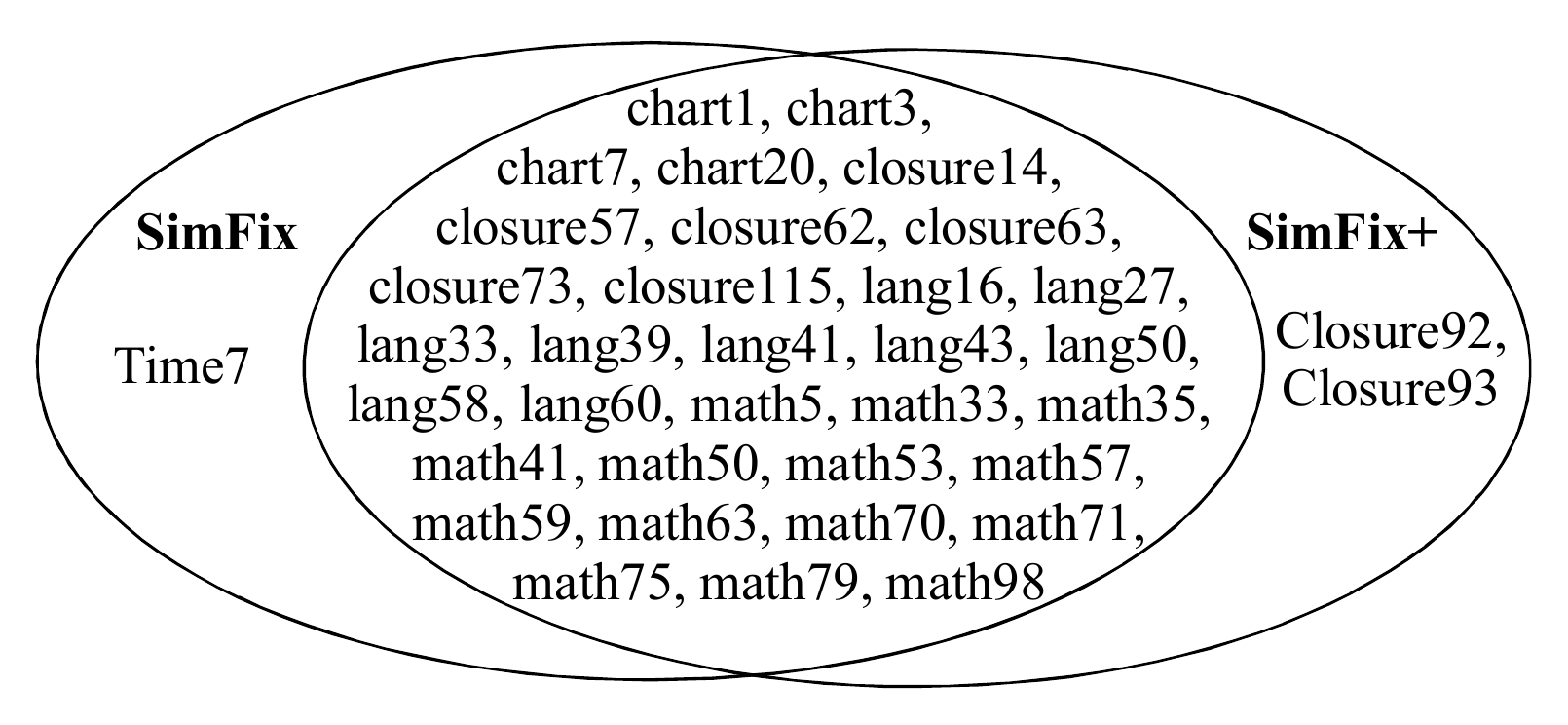}
	\caption{Faults in \dfj bugs for which SimFix and SimFix+ can build correct fixes.}\label{fig:venn}
 \end{figure}

As shown in \autoref{fig:venn},
both SimFix and SimFix+ can build \emph{correct} fixes
for the same 33 faults in \dfj.
SimFix can generate correct fixes for 1 other fault that SimFix+ cannot,
and hence can correctly fix 34 faults in total;
conversely, SimFix+ can generate correct fixes for another 2 faults that SimFix cannot, and hence can correctly fix 35 in total.
As in the case of the valid fixes, the differences are due to higher ranks of locations that lead to suitable donor code against lower ranks of donor code that is useful for mutation-based fault localization (or vice versa) 
in certain conditions.

\begin{table}[!tb]
   \centering
   \setlength{\tabcolsep}{1.7pt}
   \footnotesize\captionsetup{font=footnotesize}
   \caption{\footnotesize 
	Summary statistics of the experiments on SimFix and SimFix+.
	For each \textsc{measure}:
	the \emph{relative cost} $\frac{\sum \textrm{SimFix+}}{\sum \textrm{SimFix}}$ of SimFix+ over SimFix;
	the \emph{mean cost difference} $\mean(\textrm{SimFix} - \textrm{SimFix+})$ between SimFix and SimFix+;
	the estimate $\widehat{b}$ of \textsl{slope} $b$ expressing \restore's cost as a linear function of SimFix, with 95\% probability interval $(b_l, b_h)$;
	the estimate $\widehat{\chi}$ and upper bound $\chi_h$ on the \emph{crossing ratio} $\chi$.}
\label{tab:simfix-summary-results}
   \begin{tabular*}{\linewidth}{@{\extracolsep{\fill} }c c c rrr rr}
      \toprule
      \multirow{2}{*}[-4pt]{\textsc{measure}}& \multirow{2}{*}[-4pt]{$\frac{\sum \textrm{SimFix+}}{\sum \textrm{SimFix}}$} & \multirow{2}{*}[-4pt]{\scriptsize $\mean(\textrm{SimFix} - \textrm{SimFix+})$} & \multicolumn{3}{c}{\textsl{slope} $b$: 95\%}
   & \multicolumn{2}{c}{\textsl{crossing} $\chi$}
   \\
     \cmidrule(lr){4-6}
     \cmidrule(lr){7-8}
    & 
    & 
    & $b_l$ & $\widehat{b}$ & $b_h$
    & $\widehat{\chi}$ & $\chi_h$
   \\ \midrule
   \textsc{t2v}  & 0.69 & \makebox[1cm][r]{14} & 0.5 & 0.6 & 0.7 & 0.03 & 0.15
\\
\textsc{t2c}  & 0.63 & \makebox[1cm][r]{9} & 0.3 & 0.5 & 0.6 & 0.06 & 0.20
\\
\textsc{c2v}  & 0.60 & \makebox[1cm][r]{238} & 0.4 & 0.5 & 0.6 & 0.02 & 0.08
\\
\textsc{c2c}  & 0.55 & \makebox[1cm][r]{166} & 0.3 & 0.5 & 0.7 & 0.01 & 0.16
\\

   \bottomrule
   \end{tabular*}
   \end{table}

\begin{figure}[!tb]
	\centering\captionsetup[subfigure]{font=footnotesize}\captionsetup{font=footnotesize}
	\begin{subfigure}[t]{0.2\textwidth}
		\includegraphics[width=\textwidth]{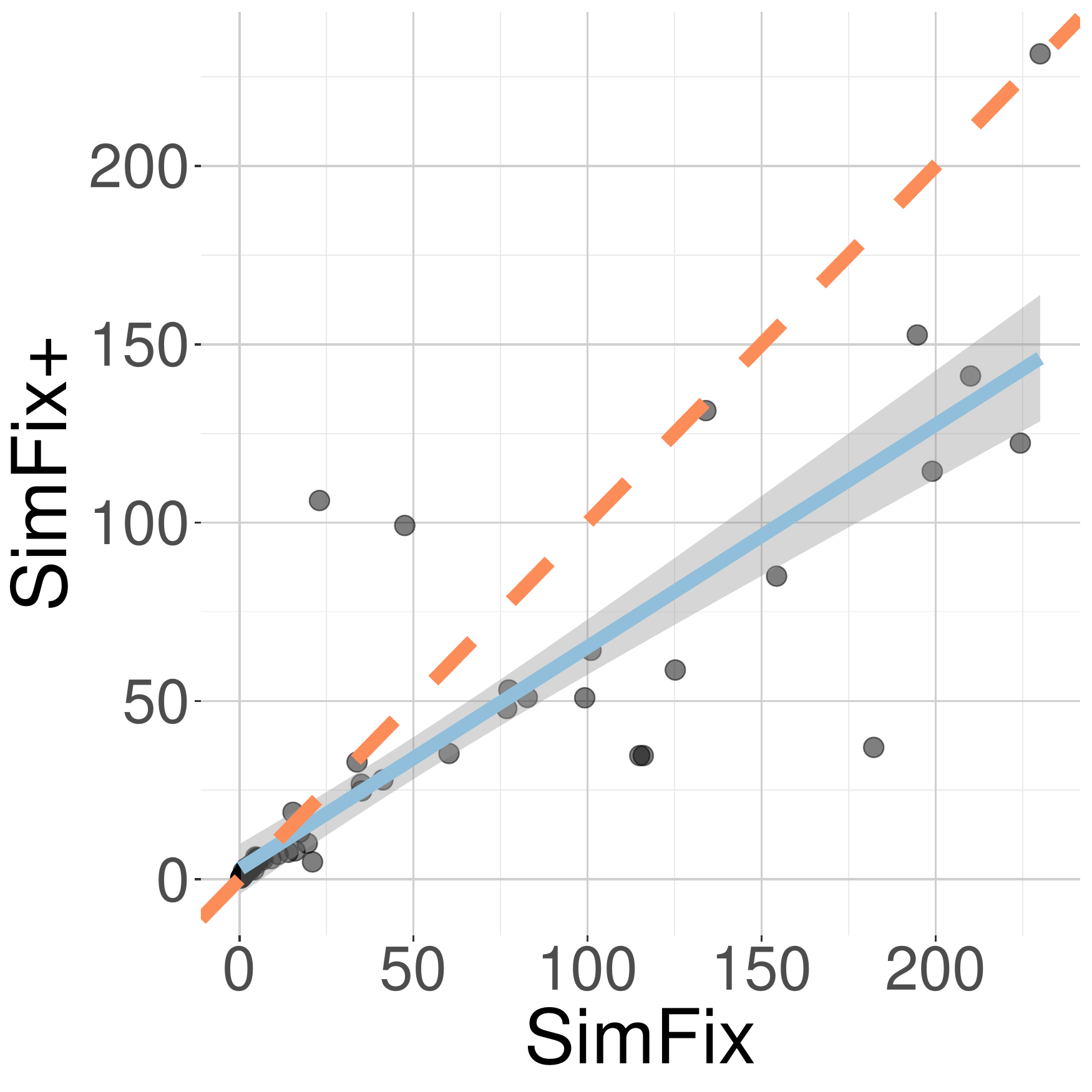}
		\caption{\textsc{t2v}}
		\label{fig:simfix-valid-t}
   \end{subfigure}
   \hfil
	\centering\captionsetup[subfigure]{font=footnotesize}\captionsetup{font=footnotesize}
	\begin{subfigure}[t]{0.2\textwidth}
		\includegraphics[width=\textwidth]{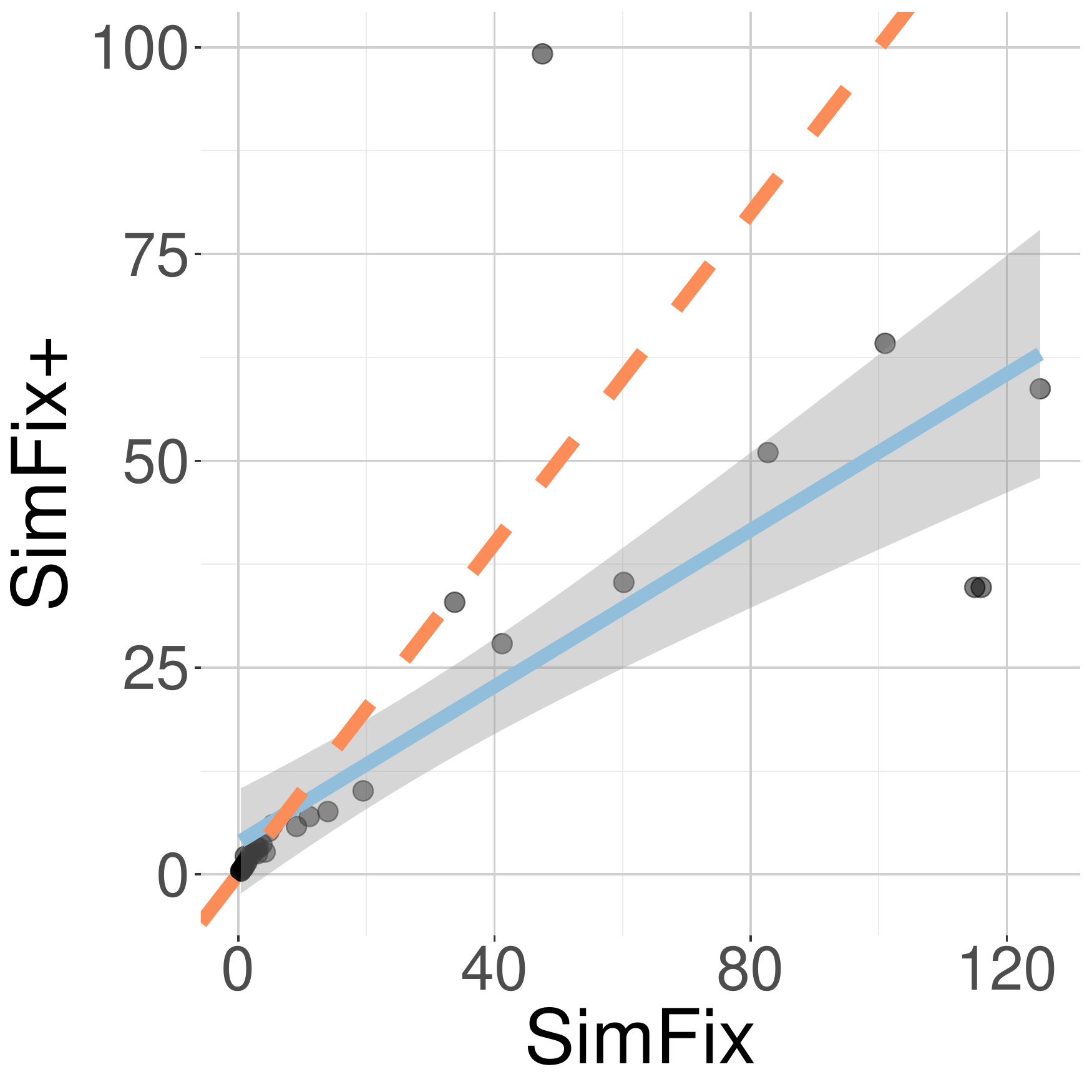}
		\caption{\textsc{t2c}}
		\label{fig:simfix-correct-t}
	\end{subfigure}
   
	\begin{subfigure}[t]{0.2\textwidth}
		\includegraphics[width=\textwidth]{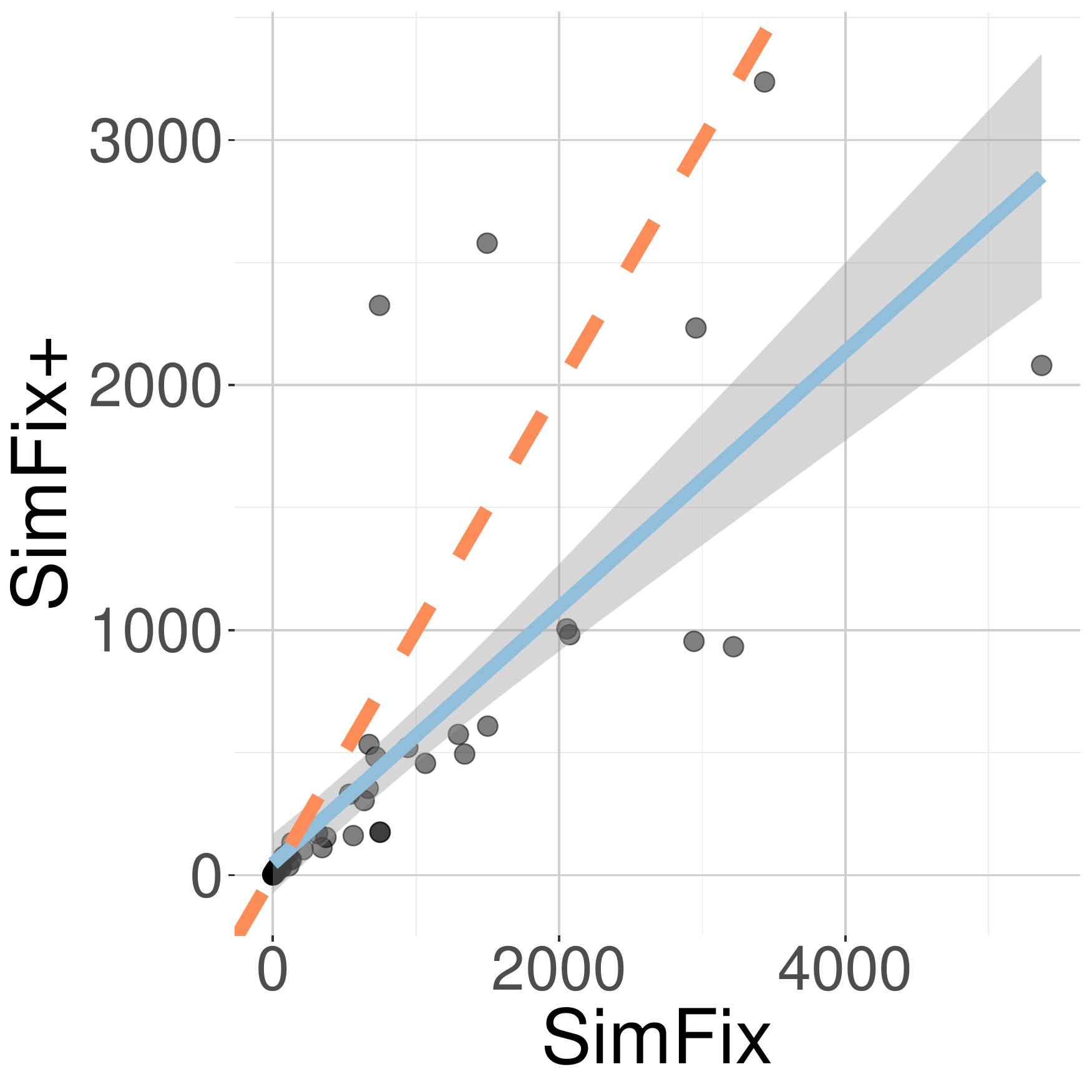}
		\caption{\textsc{c2v}}
		\label{fig:simfix-valid-c2v}
   \end{subfigure}  
   \hfil
   \begin{subfigure}[t]{0.2\textwidth}
		\includegraphics[width=\textwidth]{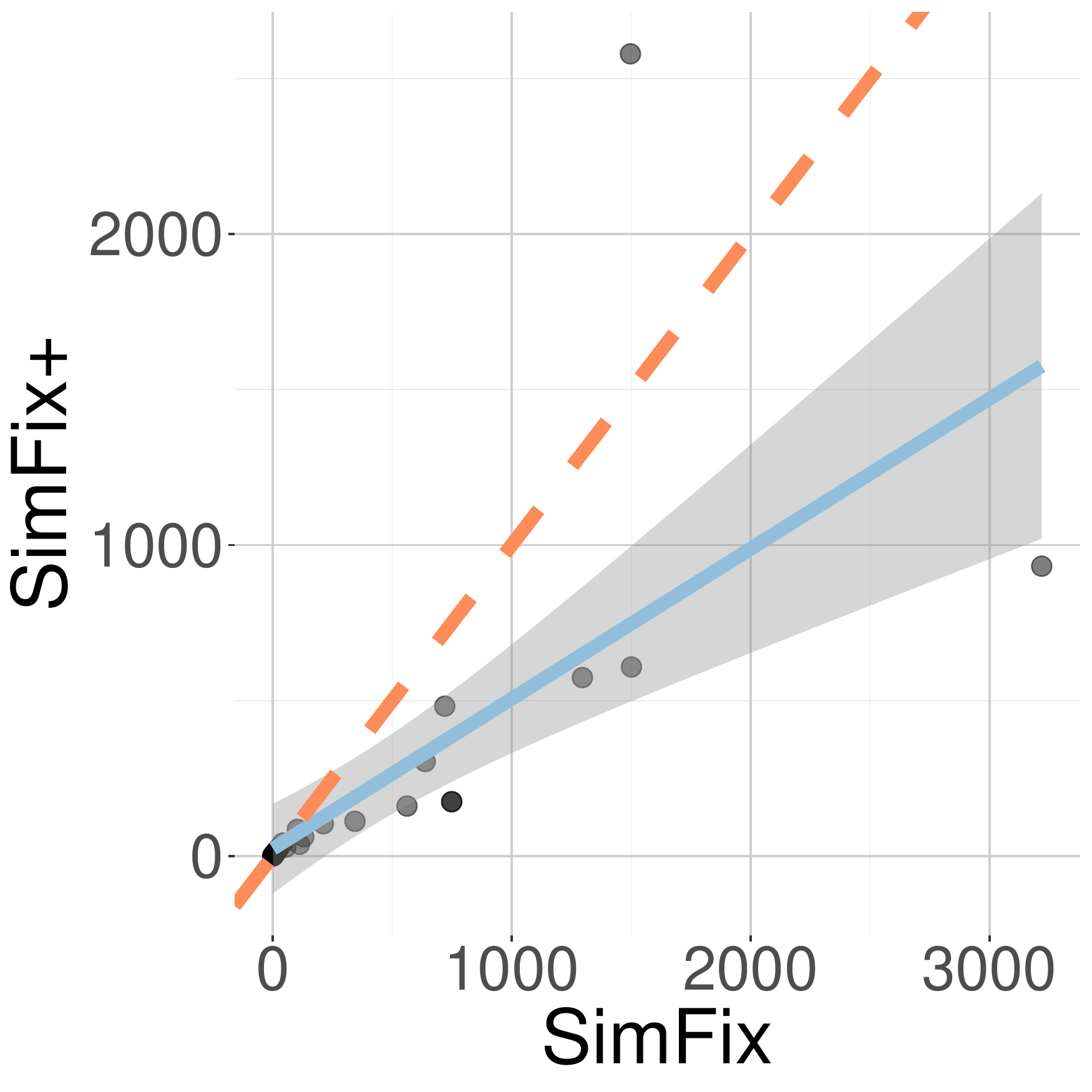}
		\caption{\textsc{c2c}}
		\label{fig:simfix-correct-c2c}
	\end{subfigure}  
	\caption{Comparison of SimFix and SimFix+ on various measures.
     For each measure $m$, a point with coordinates $x = u, y = v$
     indicates that SimFix costed $u$ on a certain fault while SimFix+ costed $v$ on the same fault.
   As in \autoref{fig:linregs}, the {\color{cred}dashed line} is $y = x$; 
the {\color{cblue}solid line} is the linear regression with $y$ dependent on $x$.}
	\label{fig:simfix-comparison-correct}
\end{figure}

How does SimFix+ compares to SimFix on the \emph{large majority} of \dfj faults
where both tools are successful?
For the 64 \dfj faults that both can repair with at least a \emph{valid} fix,
\autoref{fig:simfix-valid-t} and \autoref{fig:simfix-valid-c2v}
visually compare total running time (\textsc{t2v})\footnote{Since SimFix and SimFix+ stop after one valid fix is built, total running time \textsc{t} and running time \textsc{t2v} until a valid fix is found coincide.} and number of candidates checked (\textsc{c2v}) until a valid fix is found.
When both SimFix and SimFix+ are successful, the latter is decidedly more \emph{efficient}: 
the summary statistics of \autoref{tab:simfix-summary-results}
confirm that 
it takes 69\% of the running time, and needs to check 60\% as many candidates.
For the 33 \dfj faults that both tools can repair with a \emph{correct} fix,
the advantage of SimFix+ over SimFix
in terms of total running time (\textsc{t2c})
and number of candidates checked (\textsc{c2c}) until a correct fix is found 
is also evident, as shown in \autoref{fig:simfix-valid-t},  \autoref{fig:simfix-valid-c2v}, and \autoref{tab:simfix-summary-results}.

Unlike \restore---which ``uses'' some of the efficiency brought by retrospective fault localization to explore a larger fix space than \jaid---SimFix+
has exactly the same fix space as SimFix.
What we found in this section's experiments is consistent with this design choice:
SimFix+ has an effectiveness that is very similar to that of SimFix (precisely, slightly better precision and recall);
retrospective fault localization brings clear improvements
but mostly in terms of efficiency.
Trading off some of this greater efficiency to explore a larger fix space belongs to future work.

\begin{result}
  Retrospective fault localization implemented atop SimFix
  cuts down the running time of the tool by 30\% or more,\\
  without negatively affecting bug-fixing effectiveness.
\end{result}

\subsection{Threats to Validity}

\nicepar{Construct validity.} Threats to construct validity are concerned with whether 
the measurements taken in the evaluation realistically capture the phenomena under investigation.

An important measure is the number of \emph{correct} fixes---fixes that are semantically equivalent to programmer-written fixes for the same fault. 
Since correctness is manually assessed, 
different programmers may disagree with the authors' classifications in some cases. 
To mitigate the threat, we follow the common approach~\cite{DBLP:journals/ese/MartinezDSXM17,chen_contract_2017} of being conservative: fixes that do not clearly have the same behavior as the programmer-written ones 
are regarded as \emph{incorrect}.

Several measures could be used to assess the performance of automated program repair tools.
In our evaluation, we focus on measures that have a clear impact on \emph{practical usability}---especially number of valid and correct fixes, and running time.

When, in \autoref{sec:rq3}, we zoom in to analyze the behavior of different aspects of \restore's fault localization technique, we use the number of fixes generated and validated until the first valid fix is found.
This measure has been used by other evaluations of fault localization in program repair~\cite{Qi2013} 
because it assesses the overall effectiveness of fault localization in guiding the search for valid fixes---instead of measures, such as the rank of program locations, 
narrowly focused on the standard output of fault localization without context~\cite{Parnin_2011}.

Our summary statistics in \autoref{tab:summary-results} follow recommended practices~\cite{benchmarking};
in particular, we used statistics that are easy to interpret,
and based statistical significance on whether ``an estimate is at least two standard errors away from some [...] value that would indicate no effect present''~\cite{bsp}.

\textbf{Internal validity.} Threats to internal validity are mainly concerned with 
factors that may affect the evaluation results but were not properly controlled for.

One obvious threat to internal validity are possible bugs in the implementation of \tech,
or in the scripts we used to run our experiments. 
To address this threat, we reviewed our code and our experimental infrastructure between authors,
to slash chances that major errors affected the soundness of our results.

Another possible threat comes from comparing \restore to tools other than \jaid based on the data of their published experimental evaluations---without \emph{repeating} the experiments on the same system used to run \restore.
This threat has only limited impact: we do not compare \restore to tools other than \jaid on measures of performance---which require a uniform runtime environment---but only on measures of effectiveness such as precision and recall---which record each tool's bug-fixing capabilities on the same \dfj benchmark.

\textbf{External validity.} Threats to external validity are mainly concerned with whether 
our findings generalize---supporting broader conclusions.

\dfj has become accepted as an effective benchmark to evaluate dynamic analysis and repair tools for Java, because of the variety and size of its curated collection of faults.
At the same time, as with every benchmark, there is the lingering risk that
new techniques become narrowly optimized for \dfj 
without ascertaining that they do not overfit the benchmark.
As future work, we plan to carry out evaluations on faults from different sources,
to strengthen our claims of external validity.

Both the implementation and the evaluation of \tech are based on the \jaid repair system, and hence the fine-grained evaluation of \restore focused on how it 
improves over \jaid.
To demonstrate that most of the ideas behind retrospective fault localization (\autoref{sec:details})
are applicable to other generate-and-validate automated program repair techniques,
we also implemented retrospective fault localization on top of SimFix~\cite{JiangXZGC18}---another
state-of-the-art program repair technique for Java.
Generalizing retrospective fault localization to work with repair techniques
that are even more different---for example, based on synthesis---belongs
to future work.

\section{Related Work}~\label{sec:relatedWork}

Research in automated program repair has gained significant traction in the decade since the publication of the first works in this area~\cite{arcuri:novel:2008,weimer2009}---often
taking advantage of advances in fault localization.
In this section, we focus on reviewing the approaches that have more directly influenced the design of \restore.
Other publications provide comprehensive summaries of fault localization~\cite{Wong2016} and automated program repair~\cite{Monperrus2017,survey2-apr} techniques.

\subsection{Fault Localization}
\label{sec:related-fl}

The goal of fault localization is finding positions in the source code of a faulty program that are responsible for the fault.
The concrete output of a fault localization technique is a list of statements, branches, or program states ranked according to their likelihood of being implicated with a fault.
By focusing their attention on specific parts of a faulty program, 
such lists should help programmers debugging and patching.
While this information may not be enough for human programmers~\cite{Parnin_2011},
it is a fundamental ingredient of \emph{automated} program repair.
Thus, research in fault localization has seen a resurgence as part of an effort to improve automated repair.

\emph{Spectrum-based} fault localization techniques~\cite{Naish2011,Abreu2007} are among the most extensively studied.
The basic idea of spectrum-based fault localization is to use coverage information 
from tests to infer suspiciousness values of program entities (statements, branches, or states):
for example, a statement executed mostly by failing tests is more suspicious than one executed mostly by passing tests. 

Several automated program repair techniques use spectrum-based fault localization algorithms~\cite{weimer2009,Debroy2010,nguyen2013,Kim2013,pei_automated_2014,chen_contract_2017}.
Generating a correct fix, however, typically requires more information than the suspiciousness ranking provided by spectrum-based techniques: 
an empirical evaluation of 15 popular spectrum-based fault localization 
techniques~\cite{Qi2013} 
found that the typical evaluation criteria used in fault-localization research 
(namely, the suspiciousness ranking)
are not good predictors of whether a technique will perform well in automated program repair.
This observation buttresses our suggestion that fault localization should
be \emph{co-designed} with automated program repair to perform better---as we did with
retrospective fault localization.

Fault localization needs sources of additional information to be more accurate.
One effective idea---pioneered by delta debugging~\cite{Zeller2005}---is to \emph{modify} a program and observe how small local modifications 
affect its behavior in passing vs.\ failing runs. 
More recently, ideas from mutation testing~\cite{Jia2011} and delta-debugging have been combined
to perform \emph{mutation-based} fault localization:
randomly mutate a faulty program, and assess whether the mutation changes the behavior 
on passing or failing tests.

Metallaxis~\cite{papadakis_metallaxis_2015} and MUSE~\cite{Moon2014,Hong2015} 
are two representative mu\-ta\-tion-based fault localization techniques. 
Experiments with these tools indicate
that mutation-based fault localization often outperforms spectrum-based fault localization 
in different conditions~\cite{Moon2014,papadakis_metallaxis_2015}.
In our work, we used a variant of the Metallaxis algorithm, 
because it tends to perform better than MUSE with tasks similar to those we need for automated program repair.
The main downside of mutation-based fault localization is that it can be a performance hog,
because it requires to rerun tests on a large amount of mutants.
Thus, a key idea of our retrospective fault localization is to reuse, as much as possible,
validation results (which have to be performed anyway for program repair) 
to perform mutation-based analysis.

In retrospective fault localization,
a simple fault-localization process bootstraps a feedback loop
that implements a more accurate mutation-based fault localization.
\restore currently uses a spectrum-based technique for the bootstrap phase (see \autoref{sec:traditional-fault-localization});
however, other fault localization techniques---such as those based on
statistical analysis~\cite{Liblit_2005,Liu2006sdh}, machine learning~\cite{Briand2007UML,WongQ09bnn}, or deep learning~\cite{gupta2019deep}---could be used instead.
Even techniques that are not designed specifically for fault localization may be used, as long as they produce a ranked list of suspicious program entities.
For example, MintHint~\cite{Kaleeswaran2014MAS} performs a correlation analysis to
identify expressions that should be changed to fix faults.
The expressions, or more generally their program locations,
could thus be treated as suspicious entities for the purpose of initiating fault localization.

\subsection{Automated Program Repair}
\label{sec:related-apr}

\nicepar{Generate-and-validate} (G\&V)
remains the most widespread approach to automated program repair:
given a faulty program and a group of passing and failing tests, 
generate fix candidates by heuristically searching a program space;
then, check the validity of candidates by rerunning all available tests.
GenProg~\cite{weimer2009,goues_systematic_2012} pioneered G\&V repair
by using genetic programming to mutate a faulty program and generate fix candidates.
RERepair~\cite{Qi2014} works similarly to GenProg but uses random search instead of genetic programming.
AE~\cite{weimer_leveraging_2013} enumerates variants systematically, and uses simple semantic checks to reduce the number of equivalent fix candidates that have to be validated.
Par~\cite{Kim2013} uses patterns modeled after existing programmer-written fixes to 
guide the search toward generating fixes that are easier for programmers to understand.

This first generation of G\&V tools is capable of working on real-world bugs,
but has the tendency to \emph{overfit} the input tests~\cite{Smith2015}---thus generating many fixes that pass validation but are not actually correct~\cite{Qi_2015}.
A newer generation of tools addressed this shortcoming by supplying G\&V program repair with \emph{additional information}, often coming from mining human-written fixes.
AutoFix~\cite{pei_automated_2014} uses contracts (assertions such as pre- and postconditions) 
to improve the accuracy of fault localization.
SPR~\cite{long_staged_2015} generates candidate fixes according to a set of predefined transformation functions;
Prophet~\cite{long_2016} implements a probabilistic model, learned by mining human-written patches, on top of SPR to direct the search towards fixes with a higher chance of being correct.
HDA~\cite{xuan_2016} performs a stochastic search similar to genetic programming,
and uses heuristics mined from fix histories available in public bug repositories to guide
the search toward generating correct fixes.
ACS~\cite{Xiong_2017} builds precise changes of conditional predicates, 
based on a combination of dependency analysis and mining API documentations.
Genesis~\cite{genesis} learns templates for code transformations from human patches, 
and instantiates the templates to generate new fixes.
ssFix~\cite{Xin2017} matches contextual information at the fixing location to a database of human-written fixes, and uses this to drive fix generation.
\jaid~\cite{chen_contract_2017} uses rich state abstractions in fault localization to
generate correct repairs for a variety of bugs.
Elixir~\cite{Saha2017} specializes in repairing buggy method invocations, using machine-learned models to 
prioritize the most effective repairs.
SimFix~\cite{JiangXZGC18} combines the information extracted from existing patches and snippets 
similar to the code under fix to make the search for correct fixes more efficient. 
CapGen~\cite{WenCWHC18} improves the effectiveness of expression-level fix generation by leveraging fault context information 
so that fixes more likely to be correct are generated first.
SketchFix~\cite{HuaZWK18} expresses 
program repair as a sketching problem~\cite{sketching} with
``holes'' in suspicious statements,
and uses synthesis to fill in the holes with plausible replacements.
\restore and SketchFix both 
work to better integrate phases that are normally separate in automated repair---fault localization and fix validation in \restore,
and fix generation and fix validation in SketchFix.

Most of these tools are quite effective at generating correct fixes for real bugs; 
several of them do so by mining \emph{additional information}. 
Further improvements in G\&V repair hinge on the capability of improving the precision of fault localization.
A promising option is using mutation-based fault localization,
which was recently investigated~\cite{timperley_investigation_2017} 
on data from the BugZoo\footnote{\url{https://github.com/squaresLab/BugZoo}} repair benchmarks.
\cite{timperley_investigation_2017} found no significant improvement 
on the overall repair performance---supposedly 
because the single-edit mutations used in the study may be too simple 
to reveal substantial differences between programs variants.

In our retrospective fault localization, we combine mutation testing 
with a G\&V technique that can generate complex ``higher-order'' program mutants,
and tightly integrate fault localization and fix generation.
This way, \restore benefits from the additional accuracy of mutation-based fault localization
without incurring the major overhead typical of mutation testing.

\nicepar{Test selection and prioritization} has been studied in the context of G\&V automated program repair to improve the efficiency of fix evaluation.
For example, techniques based on genetic programming---such
as GenProg~\cite{weimer2009} and PAR~\cite{Kim2013}---can become very computationally expensive if they evaluate all program mutations on all available tests.
To improve this situation,
one could use all the failing tests but only a small sample of the passing tests---selected
randomly~\cite{Fast10dbf} or 
using an adaptive test suite reduction strategy~\cite{Walcott2006TTS}.
Another approach is the FRTP technique~\cite{Qi13eapr,Qi2014},
which gives higher priority to a test
the more fixes it has invalidated in previous iterations.
\restore currently uses a very simple test selection strategy for partial validation (\autoref{sec:partial-validation})
consisting in just running the originally failing tests.
This was quite economical, yet effective, in the experiments with \dfj,
but cannot replace a full validation step.
To achieve further improvements we will consider more sophisticated test selection strategies in future work.

\nicepar{Correct-by-construction} 
program repair techniques~\cite{avgustinov2015,mechtaev2016,nguyen2013,xuan2016,Le2017} 
express the repair problem as a constraint satisfaction problem,
and then use constraint solver to build fixes that satisfy those constraints.
Relying on static instead of dynamic analysis makes correct-by-construction 
techniques generally \emph{faster}
than G\&V ones, and is particularly effective when looking for fixes with a restricted, simple form.

\section{Conclusions}~\label{sec:conclusion}
We presented \emph{retrospective fault localization}:
a novel fault localization technique that integrates into 
the standard generate-and-validate process 
followed by numerous automated program repair techniques.
By executing a form of mutation-based testing using byproducts of automated repair,
retrospective fault localization delivers accurate fault localization information
while curtailing the otherwise demanding costs of running mutation-based testing.

Our experiments compared \restore---implementing retrospective fault localization---with 
13 other state-of-the-art Java program repair tools---including \jaid, 
upon which \restore's implementation is built.
They showed that \restore 
is a state-of-the-art program repair tool
that can search a large fix space---correctly 
fixing 41 faults from the \dfj benchmark, 8 that no other tool can fix---with
drastically improved performance 
(speedup over 3,  and  candidates that have to be checked cut in half).

Retrospective fault localization is a sufficiently general technique that it could be
integrated, possibly with some changes, into other generate-and-validate program repair systems.
To support this claim,
we implemented it atop SimFix~\cite{JiangXZGC18}---another recent automated program repair tool for Java---and showed it brings similar benefits in terms of improved efficiency.
As part of future work, we plan to 
combine retrospective fault localization with other recent advances in fault localization---thus furthering the exciting progress of automated program repair research.


\end{document}